# Known Algorithms on Graphs of Bounded Treewidth are Probably Optimal

Daniel Lokshtanov[*]    Dániel Marx[†]    Saket Saurabh[‡]

September 25, 2018

**Abstract**

We obtain a number of lower bounds on the running time of algorithms solving problems on graphs of bounded treewidth. We prove the results under the Strong Exponential Time Hypothesis of Impagliazzo and Paturi. In particular, assuming that SAT cannot be solved in $(2-\epsilon)^n m^{\mathcal{O}(1)}$ time, we show that for any $\epsilon > 0$;

- INDEPENDENT SET cannot be solved in $(2-\epsilon)^{\mathbf{tw}(G)}|V(G)|^{\mathcal{O}(1)}$ time,
- DOMINATING SET cannot be solved in $(3-\epsilon)^{\mathbf{tw}(G)}|V(G)|^{\mathcal{O}(1)}$ time,
- MAX CUT cannot be solved in $(2-\epsilon)^{\mathbf{tw}(G)}|V(G)|^{\mathcal{O}(1)}$ time,
- ODD CYCLE TRANSVERSAL cannot be solved in $(3-\epsilon)^{\mathbf{tw}(G)}|V(G)|^{\mathcal{O}(1)}$ time,
- For any $q \geq 3$, $q$-COLORING cannot be solved in $(q-\epsilon)^{\mathbf{tw}(G)}|V(G)|^{\mathcal{O}(1)}$ time,
- PARTITION INTO TRIANGLES cannot be solved in $(2-\epsilon)^{\mathbf{tw}(G)}|V(G)|^{\mathcal{O}(1)}$ time.

Our lower bounds match the running times for the best known algorithms for the problems, up to the $\epsilon$ in the base.

## 1 Introduction

It is well-known that many NP-hard graph problems can be solved efficiently if the *treewidth* ($\mathbf{tw}(G)$) of the input graph $G$ is bounded. For an example, an expository algorithm to solve VERTEX COVER and INDEPENDENT SET running in time $\mathcal{O}^*(4^{\mathbf{tw}(G)})$ is described in the algorithms textbook by Kleinberg and Tardos [15] (the $\mathcal{O}^*$ notation suppresses factors polynomial in the input size), while the book of Niedermeier [20] on fixed-parameter algorithms presents an algorithm with running time $\mathcal{O}^*(2^{\mathbf{tw}(G)})$. Similar algorithms, with running times on the form $\mathcal{O}^*(c^{\mathbf{tw}(G)})$ for a constant $c$, are known for many other graph problems such as DOMINATING SET, $q$-COLORING and ODD CYCLE TRANSVERSAL [1, 9, 10, 27]. Algorithms for graph problems on bounded treewidth graphs have found many uses as subroutines in approximation algorithms [7, 8], parameterized algorithms [6, 19, 26], and exact algorithms [12, 23, 28].

In this paper, we show that any improvement over the currently best known algorithms for a number of well-studied problems on graphs of bounded treewidth would yield a faster algorithm for SAT. In particular, we show if there exists an $\epsilon > 0$ such that

- INDEPENDENT SET can be solved in $\mathcal{O}^*((2-\epsilon)^{\mathbf{tw}(G)})$ time, or
- DOMINATING SET can be solved in $\mathcal{O}^*((3-\epsilon)^{\mathbf{tw}(G)})$ time, or
- MAX CUT can be solved in $\mathcal{O}^*((2-\epsilon)^{\mathbf{tw}(G)})$ time, or

---

[*]Department of Informatics, University of Bergen, Norway. daniello@ii.uib.no
[†]School of Computer Science, Tel Aviv University, Tel Aviv, Israel. dmarx@cs.bme.hu
[‡]The Institute of Mathematical Sciences, India. saket@imsc.res.in

- ODD CYCLE TRANSVERSAL can be solved in $\mathcal{O}^*((3-\epsilon)^{\mathbf{tw}(G)})$ time, or
- there is a $q \geq 3$ such that $q$-COLORING can be solved in $\mathcal{O}^*((q-\epsilon)^{\mathbf{tw}(G)})$ time, or
- PARTITION INTO TRIANGLES can be solved in $\mathcal{O}^*((2-\epsilon)^{\mathbf{tw}(G)})$ time,

then SAT can be solved in $\mathcal{O}^*((2-\delta)^n)$ time for some $\delta > 0$. Here $n$ is the number of variables in the input formula to SAT. Such an algorithm would violate the *Strong Exponential Time Hypothesis* (SETH) of Impagliazzo and Paturi [13]. Thus, assuming SETH, the known algorithms for the mentioned problems on graphs of bounded treewidth are essentially the best possible.

To show our results we give polynomial time many-one reductions that transform $n$-variable boolean formulas $\phi$ to instances of the problems in question. Such reductions are well-known, but for our results we need to carefully control the treewidth of the graphs that our reductions output. A typical reduction creates $n$ gadgets corresponding to the $n$ variables; each gadget has a small constant number of vertices. In most cases, this implies that the treewidth can be bounded by $O(n)$. However, to prove the a lower bound of the form $\mathcal{O}^*((2-\epsilon)^{\mathbf{tw}(G)})$, we need that the treewidth of the constructed graph is $(1+o(1))n$. Thus we can afford to increase the treewidth by at most one per variable. For lower bounds above $\mathcal{O}^*((2-\epsilon)^{\mathbf{tw}(G)})$, we need even more economical constructions. To understand the difficulty, consider the DOMINATING SET problem, here we want to say that if DOMINATING SET admits an algorithm with running time $\mathcal{O}^*((3-\epsilon)^{\mathbf{tw}(G)}) = \mathcal{O}^*(2^{\log(3-\epsilon)\mathbf{tw}(G)})$ for some $\epsilon > 0$, then we can solve SAT on input formulas with $n$-variables in time $\mathcal{O}^*((2-\delta)^n)$ for some $\delta > 0$. Therefore by naïvely equating the exponent in the previous sentence we get that we need to construct an instance for DOMINATING SET whose treewidth is essentially $\frac{n}{\log 3}$. In other words, each variable should increase treewidth by *less than one*. The main challenge in our reductions is to squeeze out as many combinatorial possibilities per increase of treewidth as possible. In order to control the treewidth of the graphs we construct, we upper bound the *pathwidth* ($\mathbf{pw}(G)$) of the constructed instances and use the fact that for any graph $G$, $\mathbf{tw}(G) \leq \mathbf{pw}(G)$. Thus all of our lower bounds also hold for problems on graphs of bounded pathwidth.

**Complexity Assumption:** The *Exponential Time Hypothesis* (ETH) and its strong variant (SETH) are conjectures about the exponential time complexity of $k$-SAT. The $k$-SAT problem is a restriction of SAT, where every clause in input boolean formula $\phi$ has at most $k$ literals. Let $s_k = \inf\{\delta : k\text{-SAT}$ *can be solved in* $2^{\delta n}$ *time*$\}$. The Exponential Time Hypothesis conjectured by Impagliazzo, Paturi and Zane [14] is that $s_3 > 0$. In [14] it is shown that ETH is robust, that is $s_3 > 0$ if and only if there is a $k \geq 3$ such that $s_k > 0$. In the same year it was shown that assuming ETH the sequence $\{s_k\}$ increases infinitely often [13]. Since SAT has a $\mathcal{O}^*(2^n)$ time algorithm, $\{s_k\}$ is bounded by above by one, and Impagliazzo and Paturi [13] conjecture that $1$ is indeed the limit of this sequence. In a subsequent paper [3], this conjecture is coined as SETH.

While ETH is now a widely believed assumption, and has been used as a starting point to prove running time lower bounds for numerous problems [5, 4, 11, 18, 17], SETH remains largely untouched (with one exception [21]). The reason for this is two-fold. First, the assumption that $\lim_{k \to \infty} s_k = \infty$ is a very strong one. Second, when proving lower bounds under ETH we can utilize the *Sparsification Lemma* [14] which allows us to reduce from instances of 3-SAT where the number of clauses is linear in the number of variables. Such a tool does not exist for SETH, and this seems to be a major obstruction for showing running time lower bounds for interesting problems under SETH. We overcome this obstruction by circumventing it – in order to show running time lower bounds for algorithms on bounded treewidth graphs sparsification is simply not required. We would like to stress that our results make sense, even if one does not believe in SETH. In particular, our results show that one should probably wait with trying to improve the known algorithms for graphs of bounded treewidth until a faster algorithm for SAT is around.

**Related Work.** Despite of the importance of fast algorithms on graphs of bounded treewidth or pathwidth, there is *no* known natural graph problem for which we know an algorithm outperforming the naïve approach on bounded pathwidth graphs. For treewidth, the situation is slightly better: Alber et al. [1] gave a $\mathcal{O}^*(4^{\mathbf{tw}(G)})$ time algorithm for DOMINATING SET, improving over the natural $\mathcal{O}^*(9^{\mathbf{tw}(G)})$ algorithm of Telle and Proskurowski [25]. Recently, van Rooij et al. [27] observed that one could use fast subset convolution [2] to improve the running time of algorithms on graphs of bounded treewidth. Their results include a $\mathcal{O}^*(3^{\mathbf{tw}(G)})$ algorithm for DOMINATING SET and a $\mathcal{O}^*(2^{\mathbf{tw}(G)})$ time algorithm for PARTITION INTO TRIANGLES. Interestingly, the effect of applying subset convolution was that the running time for several graph problems on bounded treewidth graphs became the same as the running time for the problems on graphs of bounded pathwidth.

In [27], van Rooij et al. believe that their algorithms are probably optimal, since the running times of their algorithms match the size of the dynamic programming table, and that improving the size of the table without losing time should be very difficult. Our results prove them right: improving their algorithm is at least as hard as giving an improved algorithm for SAT.

## 2 Preliminaries

In this section we give various definitions which we make use of in the paper. Let $G$ be a graph with vertex set $V(G)$ and edge set $E(G)$. A graph $G'$ is a *subgraph* of $G$ if $V(G') \subseteq V(G)$ and $E(G') \subseteq E(G)$. For subset $V' \subseteq V(G)$, the subgraph $G' = G[V']$ of $G$ is called a *subgraph induced by $V'$* if $E(G') = \{uv \in E(G) \mid u,v \in V'\}$. By $N(u)$ we denote (open) neighborhood of $u$ in graph $G$ that is the set of all vertices adjacent to $u$ and by $N[u] = N(u) \cup \{u\}$. Similarly, for a subset $D \subseteq V$, we define $N[D] = \cup_{v \in D} N[v]$.

A *tree decomposition* of a graph $G$ is a pair $(\mathcal{X}, T)$ where $T$ is a tree and $\mathcal{X} = \{X_i \mid i \in V(T)\}$ is a collection of subsets of $V$ such that: **1.** $\bigcup_{i \in V(T)} X_i = V(G)$, **2.** for each edge $xy \in E(G)$, $\{x, y\} \subseteq X_i$ for some $i \in V(T)$; **3.** for each $x \in V(G)$ the set $\{i \mid x \in X_i\}$ induces a connected subtree of $T$. The *width* of the tree decomposition is $\max_{i \in V(T)}\{|X_i| - 1\}$. The *treewidth* of a graph $G$ is the minimum width over all tree decompositions of $G$. We denote by $\mathbf{tw}(G)$ the treewidth of graph $G$. If in the definition of treewidth we restrict the tree $T$ to be a path then we get the notion of pathwidth and denote it by $\mathbf{pw}(G)$. For our purpose we need an equivalent definition of pathwidth via *mixed search* games.

In a mixed search game, a graph $G$ is considered as a system of tunnels. Initially, all edges are contaminated by a gas. An edge is *cleared* by placing searchers at both its end-points simultaneously or by sliding a searcher along the edge. A cleared edge is re-contaminated if there is a path from an uncleared edge to the cleared edge without any searchers on its vertices or edges. A search is a sequence of operations that can be of the following types: (a) placement of a new searcher on a vertex; (b) removal of a searcher from a vertex; (c) sliding a searcher on a vertex along an incident edge and placing the searcher on the other end. A search strategy is winning if after its termination all edges are cleared. The mixed search number of a graph G, denoted by $\mathbf{ms}(G)$, is the minimum number of searchers required for a winning strategy of mixed searching on $G$. Takahashi, Ueno and Kajitani [24] obtained the following relationship between $\mathbf{pw}(G)$ and $\mathbf{ms}(G)$, which we use for bounding the pathwidth of the graphs obtained in reduction.

**Proposition 1** ([24]). *For a graph $G$, $\mathbf{pw}(G) \leq \mathbf{ms}(G) \leq \mathbf{pw}(G) + 1$.*

An instance to SAT will always consists of a boolean formula $\phi = C_1 \wedge \cdots \wedge C_m$ over $n$ variables $\{v_1, \ldots, v_n\}$ where each clause $C_i$ is OR of one or more literals of variables. We also denote a clause $C_i$ by the set $\{\ell_1, \ell_2, \ldots, \ell_c\}$ of its literals and denote by $|C_i|$ the number of literals in $C_i$. An assignment $\tau$ to the variables is an element of $\{0,1\}^n$, and it satisfies the formula $\phi$ if for every clause $C_i$ there is literal that is assigned 1 by $\tau$. We say that a variable $v_i$ satisfies a clause $C_j$ if there exists a literal corresponding to $v_i$ in $\{\ell_1, \ell_2, \ldots, \ell_c\}$ and it is set to 1 by $\tau$. A group of variables satisfy a clause $C_j$ if there is a variable that satisfies the clause $C_j$. All the sections in this paper follows the following pattern: definition of the problem; statement of the lower bound; construction used in the reduction; correctness of the reduction; and the upper bound on the pathwidth of the resultant graph.

# 3 Independent Set

An *independent set* of a graph $G$ is a set $S \subseteq V(G)$ such that $G[S]$ contains no edges. In the INDEPENDENT SET problem we are given a graph $G$ and the objective is to find an independent set of maximum size.

**Theorem 1.** *If* INDEPENDENT SET *can be solved in* $\mathcal{O}^*((2-\epsilon)^{\mathbf{tw}(G)})$ *for some* $\epsilon > 0$ *then* SAT *can be solved in* $\mathcal{O}^*((2-\delta)^n)$ *time for some* $\delta > 0$.

**Construction.** Given an instance $\phi$ to SAT we construct a graph $G$ as follows. We assume that every clause has an even number of variables, if not we can add a single variable to all odd size clauses and force this variable to false. First we describe the construction of clause gadgets. For a clause $C = \{\ell_1, \ell_2, \ldots, \ell_c\}$ we make a gadget $\widehat{C}$ as follows. We take two paths, $CP = cp_1, cp_2 \ldots, cp_c$ and $CP' = cp'_1, cp'_2 \ldots cp'_c$ having $c$ vertices each, and connect $cp_i$ with $cp'_i$ for every $i$. For each literal $\ell_i$ we make a vertex $\ell_i$ in $\widehat{C}$ and make it adjacent to $cp_i$ and $cp'_i$. Finally we add two vertices $c_{start}$ and $c_{end}$, such that $c_{start}$ is adjacent to $cp_1$ and $c_{end}$ is adjacent to $cp_c$. Observe that the size of the maximum independent set of $\widehat{C}$ is $c+2$. Also, since $c$ is even, any independent set of size $c+2$ in $\widehat{C}$ must contain at least one vertex in $C = \{\ell_1, \ell_2, \ldots, \ell_c\}$. Finally, notice that for any $i$, there is an independent set of size $c+2$ in $\widehat{C}$ that contains $\ell_i$ and none of $\ell_j$ for $j \neq i$.

We first construct a graph $G_1$. We make $n$ paths $P_1, \ldots, P_n$, each path of length $2m$. Let the vertices of the path $P_i$ be $p_i^1 \ldots p_i^{2m}$. The path $P_i$ corresponds to the variable $v_i$. For every clause $C_i$ of $\phi$ we make a gadget $\widehat{C}_i$. Now, for every variable $v_i$, if $v_i$ occurs positively in $C_j$, we add an edge between $p_i^{2j}$ and the literal corresponding to $v_i$ in $\widehat{C}_j$. If $v_i$ occurs negatively in $C_j$, we add an edge between $p_i^{2j-1}$ and the literal corresponding to $v_i$ in $\widehat{C}_j$. Now we construct the graph $G$ as follows. We take $n+1$ copies of $G_1$, call them $G_1, \ldots G_{n+1}$. For every $i \leq n$ we connect $G_i$ and $G_{i+1}$ by connecting $p_j^{2m}$ in $G_i$ with $p_j^1$ in $G_{i+1}$ for every $j \leq n$. This concludes the construction of $G$.

**Lemma 1.** *If $\phi$ is satisfiable, then $G$ has an independent set of size $(mn + \sum_{i \leq m} |C_i| + 2)(n+1)$.*

*Proof.* Consider a satisfying assignment to $\phi$. We construct an independent set $I$ in $G$. For every variable $v_i$ if $v_i$ is set to true, then pick all the vertices on odd positions from all copies of $P_i$, that is $p_i^1, p_i^3, p_i^5$ and so on. If $v_i$ is false then pick all the vertices on even positions from all copies of $P_i$, that is $p_i^2, p_i^4, p_i^6$ and so on. It is easy to see that this is an independent set of size $mn(n+1)$ containing vertices from all the paths. We will now consider the gadget $\widehat{C}_j$ corresponding to a clause $C_j$. We will only consider the copy of $\widehat{C}_j$ in $G_1$ as the other copies can be dealt identically. Let use choose a true literal $\ell_a$ in $C_j$ and let $v_i$ be the corresponding variable. Consider the vertex $\ell_a$ in $\widehat{C}_j$. If $v_i$ occurs positively in $C_j$ then $v_i$ is true. Then $I$ does not contain $p_i^{2j}$, the only neighbour of $\ell_a$ outside of $\widehat{C}_j$. On the other hand if $v_i$ occurs negatively in $C_j$ then $v_i$ is false. In this case $I$ does not contain $p_i^{2j-1}$, the only neighbour of $\ell_a$ outside of $\widehat{C}_j$. There is an independent set of size $|C_j| + 2$ in $\widehat{C}$ that contains $\ell_a$ and none out of $\ell_b$, $b \neq a$. We add this independent set to $I$ and proceed in this manner for every clause gadget. By the end of the process $(\sum_{i \leq m} |C_i| + 2)(n+1)$ vertices from clause gadgets are added to $I$, yielding that the size of $I$ is $(mn + \sum_{i \leq m} |C_i| + 2)(n+1)$, concluding the proof. □

**Lemma 2.** *If $G$ has an independent set of size $(2mn + \sum_{i \leq m} |C_i| + 2)(n+1)$, then $\phi$ is satisfiable.*

*Proof.* Consider an independent set of $G$ of size $(mn + \sum_{i \leq m} |C_i| + 2)(n+1)$. The set $I$ can contain at most $m$ vertices from each copy of $P_i$ for every $i \leq n$ and at most $|C_j| + 2$ vertices from each copy of the gadget $C_j$. Since $I$ must contain at least these many vertices from each path and clause gadget in order to contain at least $(mn + \sum_{i \leq m} |C_i| + 2)(n+1)$ vertices, it follows that $I$ has exactly $m$ vertices in each copy of each path $P_i$ and exactly $|C_j| + 2$ vertices in each copy of each clause gadget $\widehat{C}_j$. For a fixed $j$, consider the $n+1$ copies of the path $P_j$. Since $P_j$ in $G_i$ is attached to $P_j$ in $G_{i+1}$ these $n+1$ copies of $P_i$ together form a path $P$ having $2m(n+1)$ vertices. Since $|I \cap P| = m(n+1)$ it follows that $I \cap P$ must contain every second vertex of $P$, except possibly in one position where $I \cap P$ skips two vertices of $P$. There are only $n$ paths and $n+1$ copies of $G_1$, hence the pigeon-hole principle yields that in some copy $G_y$ of $G_1$, $I$ contains every second vertex on every path $P_i$. From now onwards we only consider such a copy $G_y$.

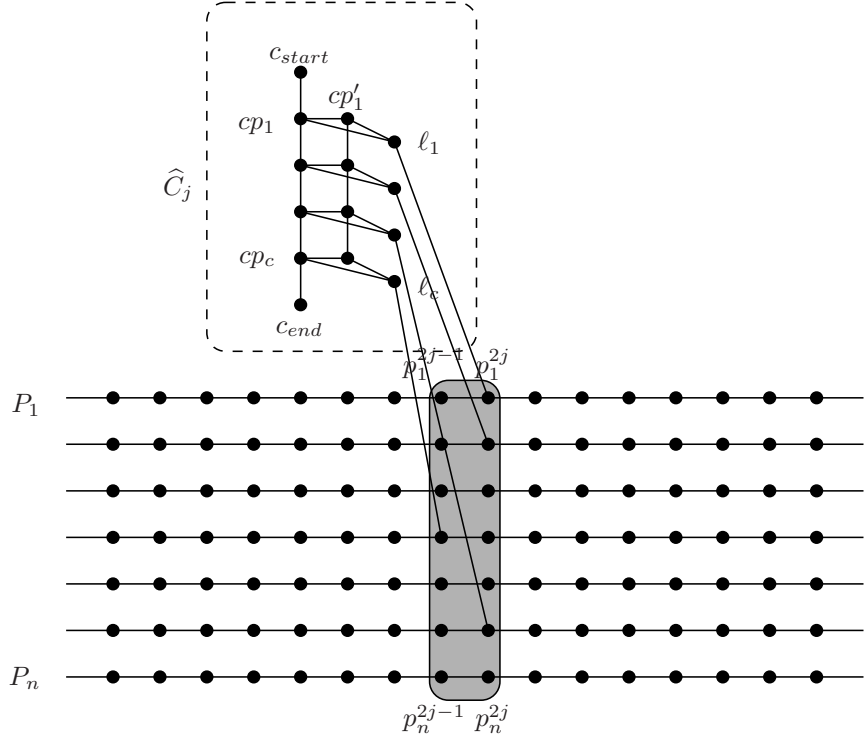

Figure 1: Reduction to INDEPENDENT SET: clause gadget $\widehat{C}_j$ attached to the $n$ paths representing the variables.

In $G_y$, for every $i \leq n$, $I$ contains every second vertex of $P_i$. We make an assignment to the variables of $\phi$ as follows. If $I$ contains all the odd numbered vertices of $P_i$ then $v_i$ is set to true, otherwise $I$ contains all the even numbered vertices of $P_i$ and $v_i$ is set to false. We argue that this assignment satisfies $\phi$. Indeed, consider any clause $C_j$, and look at the gadget $\widehat{C}_j$. We know that $I$ contains $|C_j| + 2$ vertices from $\widehat{C}_j$ and hence $I$ must contain a vertex $\ell_a$ in corresponding to a literal of $C_j$. Suppose $\ell_a$ is a literal of $v_i$. Since $I$ contains $\ell_a$, if $\ell_a$ occurs positively in $C_j$, then $I$ can not contain $p_i^{2j}$ and hence $v_i$ is true. Similarly, if $\ell_a$ occurs negatively in $C_j$ then $I$ can not contain $p_i^{2j-1}$ and hence $v_i$ is false. In both cases $v_i$ satisfies $C_j$ and hence all clauses of $\phi$ are satisfied by the assignment. □

**Lemma 3.** $\mathbf{pw}(G) \leq n + 4$.

*Proof.* We give a mixed search strategy to clean $G$ using $n+3$ searchers. For every $i$ we place a searcher on the first vertex of $P_i$ in $G_1$. The $n$ searchers slide along the paths $P_1, \ldots P_n$ in $m$ rounds. In round $j$ each searcher $i$ starts on $p_i^{2j-1}$. Then, for every variable $v_i$ that occurs positively in $C_j$, the searcher $i$ slide forward to $p_i^{2j}$. Observe that at this point there is a searcher on every neighbour of the gadget $\widehat{C}_j$. This gadget can now be cleaned with 3 additional searchers. After $\widehat{C}_j$ is clean, the additional 3 searchers are removed, and each of the $n$ searchers on the paths $P_1, \ldots P_n$ slide forward along these paths, such that searcher $i$ stands on $p_i^{2(j+1)}$. At that point, the next round commences. When the searchers have cleaned $G_1$ they slide onto the first vertex of $P_1 \ldots P_n$ in $G_2$. Then they proceed to clean $G_2, \ldots, G_{n+1}$ in the same way that $G_1$ was cleaned. Now applying Proposition 1 we get that $\mathbf{pw}(G) \leq n + 4$. □

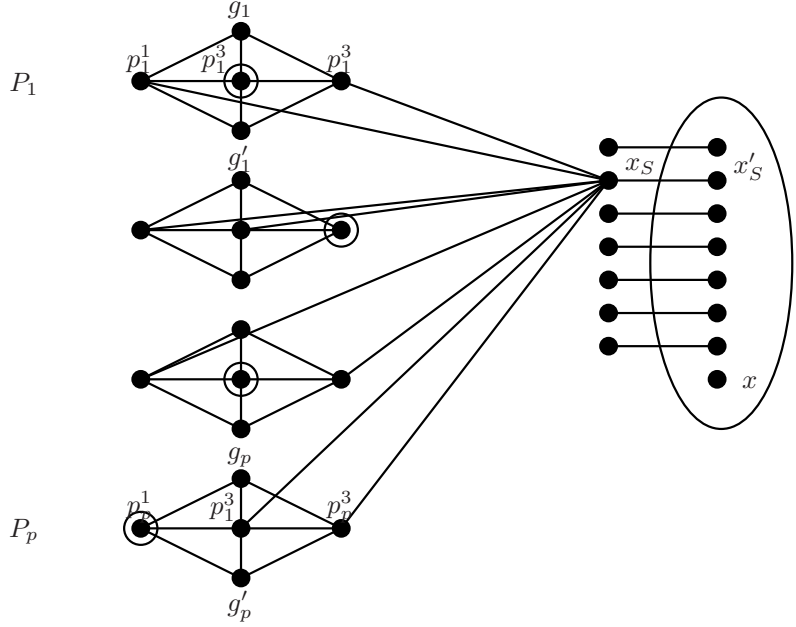

Figure 2: Reduction to DOMINATING SET: group gadget $\widehat{B}$. The set $S$ is shown by the circled vertices.

The construction, together with Lemmata 1, 2 and 3 proves Theorem 1.

## 4 Dominating Set

A *dominating set* of a graph $G$ is a set $S \subseteq V(G)$ such that $V(G) = N[S]$. In the DOMINATING SET problem we are given a graph $G$ and the objective is to find a dominating set of minimum size.

**Theorem 2.** *If* DOMINATING SET *can be solved in* $\mathcal{O}^*((3-\epsilon)^{\mathbf{pw}(G)})$ *time for some $\epsilon > 0$ then* SAT *can be solved in* $\mathcal{O}^*((2-\delta)^n)$ *time for some $\delta > 0$.*

**Construction.** Given $\epsilon < 1$ and an instance $\phi$ to SAT we construct a graph $G$ as follows. We first chose an integer $p$ depending only on $\epsilon$. Exactly how $p$ is chosen will be discussed in the proof of Theorem 2. We group the variables of $\phi$ into groups $F_1, F_2, \ldots, F_t$, each of size at most $\beta = \lfloor \log 3^p \rfloor$. Hence $t = \lceil n/\beta \rceil$. We now proceed to describe a "group gadget" $\widehat{B}$, which is central in our construction.

To build the group gadget $\widehat{B}$ we make $p$ paths $P_1, \ldots, P_p$, where the path $P_i$ contains the vertices $p_i^1$, $p_i^2$ and $p_i^3$. To each path $P_i$ we attach two *guards* $g_i$ and $g_i'$, both of which are neighbours to $p_i^1$, $p_i^2$ and $p_i^3$. When the gadgets are attached to each other, the guards will not have any neighbours outside of their own gadget $\widehat{B}$, and will ensure that at least one vertex out of $p_i^1$, $p_i^2$ and $p_i^3$ are chosen in any minimum size dominating set of $G$. Let $P$ be a vertex set containing all the vertices on the paths $P_1, \ldots, P_p$. For every subset $S$ of $P$ that picks *exactly one* vertex from each path $P_i$ we make two vertices $x_S$ and $x_S'$, where $x_S$ is adjacent to all vertices of $P \setminus S$ (all those vertices that are on paths and not in $S$) and $x_S'$ is only adjacent to $x_S$. We conclude the construction of $\widehat{B}$ by making all the vertices $x_S'$ (for every set $S$) adjacent to each other, that is making them into a clique, and adding a guard $x$ adjacent to $x_S'$ for every set $S$. Essentially $x_S'$'s together with $x$ forms a clique and all the neighbors of $x$ reside in this clique.

We construct the graph $G$ as follows. For every group $F_i$ of variables we make $m(2pt + 1)$ copies of the gadget $\widehat{B}$, call them $\widehat{B}_i^j$ for $1 \leq j \leq m(2pt+1)$. For every fixed $i \leq t$ we connect the gadgets $\widehat{B}_i^1, \widehat{B}_i^2 \ldots, \widehat{B}_i^{m(2pt+1)}$ in a path-like manner. In particular, for every $j < m(2pt+1)$ and every $\ell \leq p$ we make an edge between $p_\ell^3$ in the gadget $\widehat{B}_i^j$ with $p_\ell^1$ in the gadget $\widehat{B}_i^{j+1}$. Now we make two new vertices $h$ and $h'$, with $h$ adjacent to $h'$, $p_j^1$ in $\widehat{B}_i^1$ for every $i \leq t$, $j \leq p$ and to $p_j^3$ in $\widehat{B}_i^{m(2pt+1)}$ for every $i \leq t$, $j \leq p$. That is, for all $1 \leq i \leq t$, $h$ is adjacent to first and last vertices of "long paths" obtained after connecting the gadgets $\widehat{B}_i^1, \widehat{B}_i^2 \ldots, \widehat{B}_i^{m(2pt+1)}$ in a path-like manner.

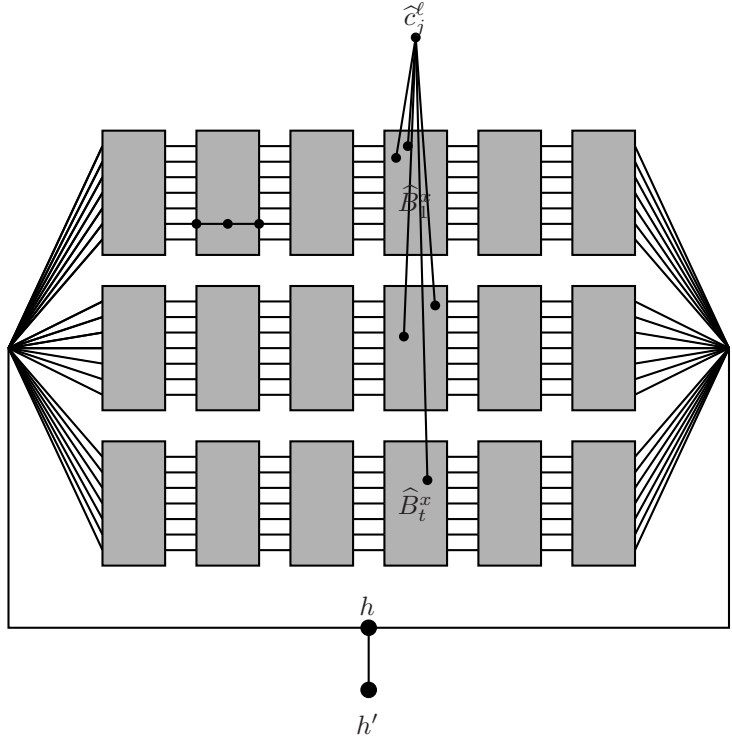

Figure 3: Reduction to DOMINATING SET: arranging the group gadgets. Note that $x = m\ell + j$, thus $\widehat{c}_j^\ell$ is attached to vertices in $\widehat{B}_1^x, \ldots, \widehat{B}_t^x$.

For every $1 \leq i \leq t$, and to every assignment of the variables in the group $F_i$, we designate a subset $S$ of $P$ in the gadget $\widehat{B}$ that picks exactly one vertex from each path $P_j$. Since there are at most $2^\beta$ different assignments to the variables in $F_i$, and there are $3^p \geq 2^\beta$ such sets $S$, we can assign a *unique* set to each assignment. Of course, the same set $S$ can correspond to one assignment of the group $F_1$ and some another assignment of the group $F_2$. Recall that the clauses of $\phi$ are $C_1, \ldots, C_m$. For every clause $C_j$ we make $2pt + 1$ vertices $\widehat{c}_j^\ell$, one for each $0 \leq \ell < 2pt + 1$. The vertex $\widehat{c}_j^\ell$ will be connected to the gadgets $\widehat{B}_i^{m\ell+j}$ for every $1 \leq i \leq t$. In particular, for every assignment of the variables in the group $F_i$ that satisfy the clause $C_j$, we consider the subset $S$ of $P$ that corresponds to the assignment. For every $0 \leq \ell < 2n + 1$, we make $x'_S$ in $\widehat{B}_i^{m\ell+j}$ adjacent to $\widehat{c}_j^\ell$. The best way to view this is that every clause $C_j$ has $2pt + 1$ private gadgets, $\widehat{B}_i^j, \widehat{B}_i^{m+j}, \ldots, \widehat{B}_i^{m2pt+j}$, in every group of gadgets corresponding to $F_i$'s. Now we have $2pt + 1$ vertices corresponding to the clause $C_j$, one each for *one* gadget from each group gadgets corresponding to $F_i$'s. This concludes the construction of $G$.

**Lemma 4.** *If $\phi$ has a satisfying assignment, then $G$ has a dominating set of size $(p+1)tm(2pt+1)+1$.*

*Proof.* Given a satisfying assignment to $\phi$ we construct a dominating set $D$ of $G$ that contains the vertex $h$ and *exactly* $p + 1$ vertices in each gadget $\widehat{B}_i^j$. For each group $F_i$ of variables we consider the set $S$ which corresponds to the restriction of the assignment to the variables in $F_i$. From each gadget $\widehat{B}_i^j$ we add the set $S$ to $D$ and also the vertex $x'_S$ to $D$. It remains to argue that $D$ is indeed a dominating set. Clearly the size is bounded by $(p + 1)tm(2pt + 1) + 1$, as the number of gadgets is $tm(2pt + 1)$.

For a fixed $i \leq t$ and $j$ consider the vertices on the path $P_j$ in the gadgets $\widehat{B}_i^\ell$ for every $\ell \leq m(2pt + 1)$. Together these vertices form a path of length $3m(2pt + 1)$ and every third vertex of this path is in $S$. Thus, all vertices on this path are dominated by other vertices on the path, except for the first and last one. Both these vertices, however, are dominated by $h$.

Now, fix some $i \leq t$ and $l \leq m(2pt+1)$ and consider the gadget $\widehat{B}_i^\ell$. Since $D$ contains some vertex on the path $P_j$, we have that for every $j$ both $g_j$ and $g'_j$ are dominated. Furthermore, for every set $S^*$ not equal to $S$ that picks exactly one vertex from each $P_j$, vertex $x_{S^*}$ is dominated by some vertex on some $P_j$—namely by all vertices in $S \setminus S^* \neq \emptyset$. The last assertion follows since $x_{S^*}$ is connected to all the vertices on paths except $S^*$. On the other hand, $x_S$ is dominated by $x'_S$, and $x'_S$ also dominates all the other vertices $x'_{S^*}$ for $S^* \neq S$ and the guard $x$.

The only vertices not yet accounted for are the vertices $\widehat{c}_j^\ell$ for every $j \leq m$ and $\ell < 2pt+1$. Fix a $j$ and a $\ell$ and consider the clause $C_j$. This clause contains a literal set to true, and this literal corresponds to a variable in the group $F_i$ for some $i \leq t$. Of course, the assignment to $F_i$ satisfies $C_j$. Let $S$ be the set corresponding to this assignment of $F_i$. By the construction of $D$, the dominating set contains $x'_S$ in $\widehat{B}_i^{m\ell+j}$ and $x'_S$ is adjacent to $\widehat{c}_j^\ell$. This concludes the proof. □

**Lemma 5.** *If $G$ has a dominating set of size $(p+1)tm(2pt+1)+1$, then $\phi$ has a satisfying assignment.*

*Proof.* Let $D$ be a dominating set of $G$ of size at most $(p+1)tm(2pt+1)+1$. Since $D$ must dominate $h'$, hence without loss of generality we can assume that $D$ contains $h$. Furthermore, inside every gadget $\widehat{B}_i^\ell$, $D$ must dominate all the guards, namely $g_j$ and $g'_j$ for every $j \leq p$, and also $x$. Thus $D$ contains at least $p+1$ vertices from each gadget $\widehat{B}_i^\ell$ which in turn implies that $D$ contains exactly $p+1$ vertices from each gadget $\widehat{B}_i^\ell$. The only way $D$ can dominate $g_j$ and $g'_j$ for every $j$ and in addition dominate $x$ with only $p+1$ verticesis if $D$ has one vertex from each $P_j$, $j \leq p$ and in addition contains some vertex in $N[x]$. Let $S$ be $D \cap P$ in $\widehat{B}_i^\ell$. Observe that $x_S$ is not dominated by $D \cap S$. The only vertex in $N[x]$ that dominates $x_S$ is $x'_S$ and hence $D$ contains $x'_S$.

Now we want to show that for every $1 \leq i \leq t$ there exists *one* $0 \leq \ell \leq 2tp$ such that for fixed $i$, $D \cap P$ is same in all the gadgets $\widehat{B}_i^{m\ell+r}$, $1 \leq r \leq m$. Consider a gadget $\widehat{B}_i^\ell$ and its follower, $\widehat{B}_i^{\ell+1}$. Let $S$ be $D \cap P$ in $\widehat{B}_i^\ell$ and $S'$ be $D \cap P$ in $\widehat{B}_i^{\ell+1}$. Observe that if $S$ contains $p_j^a$ in $\widehat{B}_i^\ell$ and $p_j^b$ in $\widehat{B}_i^{\ell+1}$ then we must have $b \leq a$. We call a consecutive pair *bad* if for some $j \leq p$, $D$ contains $p_j^a$ in $\widehat{B}_i^\ell$ and $p_j^b$ in $\widehat{B}_i^{\ell+1}$ and $b < a$. Hence for a fixed $i$, we can at most have $2p$ consecutive bad pairs. Now we mark all the bad pairs that occur among the gadgets corresponding to some $F_i$. This way we can mark only $2tp$ bad pairs. Thus, by the pigeon hole principle, there exists an $\ell \in \{0, \ldots, 2tp\}$ such that there are no bad pairs in $\widehat{B}_i^{m\ell+r}$ for all $1 \leq i \leq t$ and $1 \leq r \leq m$.

We make an assignment $\phi$ by reading off $D \cap P$ in each gadget $\widehat{B}_i^{m\ell+1}$. In particular, for every group $F_i$, we consider $S = D \cap P$ in the gadget $\widehat{B}_i^{m\ell+1}$. This set $S$ corresponds to an assignment of $F_i$, and this is the assignment of $F_i$ that we use. It remains to argue that every clause $C_r$ is satisfied by this assignment.

Consider the vertex $\widehat{c}_\ell^r$. We know that it is dominated by some $x'_S$ in a gadget $\widehat{B}_i^{m\ell+r}$. The set $S$ corresponds to an assignment of $F_i$ that satisfies the clause $C_r$. Because $D \cap P$ remains unchanged in all gadgets from $\widehat{B}_i^{m\ell+1}$ to $\widehat{B}_i^{m\ell+r}$, this is exactly the assignment $\phi$ restricted to the group $F_i$. This concludes the proof. □

**Lemma 6.** $\mathbf{pw}(G) \leq tp + \mathcal{O}(3^p)$

*Proof.* We give a mixed search strategy to clean the graph with $tp + \mathcal{O}(3^p)$ searchers. For a gadget $\widehat{B}$ we call the vertices $p_j^1$ and $p_j^3$, $1 \leq j \leq p$, as *entry vertices* and *exit vertices* respectively. We search the graph in $m(2tp+1)$ rounds. In the beginning of round $\ell$ there are searchers on the entry vertices of the gadgets $\widehat{B}_i^\ell$ for every $i \leq t$. Let $1 \leq a \leq m$ and $0 \leq b < 2tp+1$ be integers such that $\ell = a + mb$. We place a searcher on $\widehat{c}_a^b$. Then, for each $i$ between 1 and $p$ in turn we first put searchers on all vertices of $\widehat{B}_i^\ell$ and then remove all the searchers from $\widehat{B}_i^\ell$ except for the ones standing on the exit vertices. After all gadgets $\widehat{B}_1^\ell \ldots \widehat{B}_t^\ell$ have been cleaned in this manner, we can remove the searcher from $\widehat{c}_a^b$. To commence the next round, the searchers slide from the exit positions of $\widehat{B}_i^\ell$ to the entry positions of $\widehat{B}_i^{\ell+1}$ for every $i$. In total, at most $tp + |V(\widehat{B})| + 1 \leq tp + \mathcal{O}(3^p)$ searchers are used simultaneously. This together with Proposition 1 give the desired upperbound on the pathwidth. □

*Proof (of Theorem 2).* Suppose DOMINATING SET can be solved in $\mathcal{O}^*((3-\epsilon)^{\mathbf{pw}(G)}) = \mathcal{O}^*(3^{\lambda \mathbf{pw}(G)})$ time, where $\lambda = \log_3(3-\epsilon) < 1$. We choose $p$ large enough such that $\lambda \cdot \frac{p}{\lceil p \log 3 \rceil} = \frac{\delta'}{\log 3}$ for some $\delta' < 1$. Given an instance of SAT we construct an instance of DOMINATING SET using the above construction and the chosen value of $p$. Then we solve the DOMINATING SET instance using the $\mathcal{O}^*(3^{\lambda \mathbf{pw}(G)})$ time algorithm. Correctness is ensured by Lemmata 4 and 5. Lemma 6 yields that the total time taken is upper bounded by $\mathcal{O}^*(3^{\lambda \mathbf{pw}(G)}) \leq \mathcal{O}^*(3^{\lambda(tp+f(\lambda))}) \leq \mathcal{O}^*(3^{\lambda \frac{np}{\lceil p \log 3 \rceil}}) \leq \mathcal{O}^*(3^{\delta' \frac{n}{\log 3}}) \leq \mathcal{O}^*(2^{\delta'' n}) = \mathcal{O}^*((2-\delta)^n)$, for some $\delta'', \delta < 1$. This concludes the proof. □

## 5 Max Cut

A *cut* in a graph $G$ is a partition of $V(G)$ into $V_0$ and $V_1$. The *cut-set* of the cut is the set of edges whose one end point is in $V_0$ and the other in $V_1$. We say that an edge is *crossing* this cut if it has one endpoint in $V_0$ and one in $V_1$, that is, the edge is in the cut-set. The *size* of the cut is the number of edges in $G$ which are crossing this cut. If the edges of $G$ have positive integer weights then the *weight* of the cut is the sum of the weights of edges which are crossing the cut. In the MAX CUT problem we are given a graph $G$ together with an integer $t$ and asked whether there is a cut of $G$ of size at least $t$. In the WEIGHTED MAX CUT problem every edge has a positive integer weight and the objective is to find a cut of weight at least $t$.

**Theorem 3.** *If* MAX CUT *can be solved in* $\mathcal{O}^*((2-\epsilon)^{\mathbf{pw}(G)})$ *for some* $\epsilon > 0$ *then* SAT *can be solved in* $\mathcal{O}^*((2-\delta)^n)$ *time for some* $\delta > 0$.

**Construction.** Given an instance $\phi$ of SAT we first construct an instance $G_w$ of WEIGHTED MAX CUT as follows. We later explain how to obtain an instance of unweighted MAX CUT from here.

We start with making a vertex $x_0$. Without loss of generality, we will assume that $x_0 \in V_0$ in every solution. We make a vertex $\widehat{v}_i$ for each variable $v_i$. For every clause $C_j$ we make a gadget as follows. We make a path $\widehat{P}_j$ having $4|C_j|$ vertices. All the edges on $\widehat{P}_j$ have weight $3n$. Now, we make the first and last vertex of $\widehat{P}_j$ adjacent to $x_0$ with an edge of weight $3n$. Thus the path $\widehat{P}_j$ plus the edges from the first and last vertex of $\widehat{P}_j$ to $x_0$ form an odd cycle $\widehat{C}_j$. We will say that the first, third, fifth, etc, vertices are on *odd positions* on $\widehat{P}_j$ while the remaining vertices are on *even positions*. For every variable $v_i$ that appears positively in $C_j$ we select a vertex $p$ at an even position (but not the last vertex) on $\widehat{P}_j$ and make $\widehat{v}$ adjacent to $p$ and $p$'s successor on $\widehat{P}_j$ with edges of weight 1. For every variable $v_i$ that appears negatively in $C_j$ we select a vertex $p$ at an odd position on $\widehat{P}_j$ and make $\widehat{v}$ adjacent to $p$ and $p$'s successor on $\widehat{P}_j$ with edges of weight 1. We make sure that each vertex on $\widehat{P}_j$ receives an edge at most once in this process. There are more than enough vertices on $\widehat{P}_j$ to accommodate all the edges incident to vertices corresponding to variables in the clause $C_j$. We create such a gadget for each clause and set $t = 1 + (12n+1) \sum_{j=1}^{m} |C_j|$. This concludes the construction.

**Lemma 7.** *If $\phi$ is satisfiable, then $G_w$ has a cut of weight at least $t$.*

*Proof.* Suppose $\phi$ is satisfiable. We put $x_0$ in $V_0$ and for every variable $v_i$ we put $\widehat{v}_i$ in $V_1$ if $v_i$ is true and $\widehat{v}_i$ in $V_0$ if $v_i$ is false. For every clause $C_j$ we proceed as follows. Let us choose a true literal of $C_j$ and suppose that this literal corresponds to a vertex $p_j$ on $\widehat{P}_j$. We put the first vertex on $\widehat{P}_j$ in $V_1$, the second in $V_0$ and then we proceed along $\widehat{P}_j$ putting every second vertex into $V_1$ and $V_0$ until we reach $p_j$. The successor $p'_j$ of $p_j$ on $\widehat{P}_j$ is put into the same set as $p_j$. Then we continue along $\widehat{P}_j$ putting every second vertex in $V_1$ and $V_0$. Notice that even though $C_j$ may contain more than one literal that is set to true, we only select one vertex $p_j$ from the path $\widehat{P}_j$ and put $p_j$ and its successor on the same side of the partition. It remains to argue that this cut has weight at least $t$.

For every clause $C_j$ all edges on the path $\widehat{P}_j$ except for $p_j p'_j$ are crossing, and the two edges to $x_0$ from the first and last vertex of $\widehat{P}_j$ are crossing as well. These edges contribute $12n|C_j|$ to the weight of the cut. We know that $p_j$ corresponds to a literal that is set to true, and this literal corresponds to a variable $v_i$. If $v_i$ occurs positively in $C_j$ then $v_i \in V_1$ and $p_j$ is on an even position of $\widehat{P}_j$. Thus both $p_j$ and his successor $p'_j$ are in $V_0$ and hence both $v_i p_j$ and $v_i p'_j$ are crossing, contributing 2 to the weight of the cut. For each of the remaining variables $v_{i'}$ appearing in $C_j$, one of the two neighbours of $\widehat{v}_{i'}$ on $\widehat{P}_j$ appear in $V_0$ and one in $V_1$, so exactly one edge from $v_{i'}$ to $\widehat{P}_j$ is crossing. Thus the total weight of the cut is $t = \sum_{j=1}^{m} 12n|C_j| + |C_j| + 1 = m + (12n+1)\sum_{j=1}^{m}|C_j|$. This completes the proof. □

**Lemma 8.** *If $G_w$ has a cut of weight at least $t$, then $\phi$ is satisfiable.*

*Proof.* Let $(V_0, V_1)$ be a cut of $G$ of maximum weight, hence the weight of this cut is at least $t$. Without loss of generality, let $x_0 \in V_0$. For every clause $C_j$ at least one edge of the odd cycle $\widehat{C}_j$ is not crossing. If more than one edge of this cycle is not crossing, then the total weight of the cut edges incident to the path $\widehat{P}_j$ is at most $3n(4|C_j|-1) + 2n < 12|C_j|$. In this case we could change the partition $(V_0, V_1)$ such that all edges of $\widehat{P}_j$ are crossing and the first vertex of $\widehat{P}_j$ is in $V_1$. Using the new partition the weight of the crossing edges in the cycle $\widehat{C}_j$ is at least $12|C_j|$ and the edges not incident to $\widehat{P}_j$ are unaffected by the changes. This contradicts that $(V_0, V_1)$ was a maximum weight cut. Thus it follows that exactly one edge of $\widehat{C}_j$ is not crossing.

Given the cut $(V_0, V_1)$ we set each variable $v_i$ to true if $\widehat{v}_i \in V_1$ and $v_i$ to false otherwise. Consider a clause $C_j$ and a variable $v_i$ that appears in $C_j$. Let $uv$ be the edge of $\widehat{C}'_j$ that is not crossing. If there is a variable $\widehat{v}_i$ adjacent to both $u$ and $v$, then it is possible that both $\widehat{v}_i u$ and $\widehat{v}_i v$ are crossing. For every other variable $v_{i'}$ in $C_j$, at most one of the edges from $\widehat{v}_{i'}$ to $\widehat{P}_j$ is crossing. Thus, the weight of the edges that are crossing in the gadget $\widehat{C}_j$ is at most $(12n+1)|C_j| + 1$. Hence, to find a cut-set of weight at least $t$ in $G$, we need to have crossing edges in $\widehat{C}_j$ with sum of their weights exactly equal to $12n|C_j| + |C_j| + 1$. It follows that there is a vertex $\widehat{v}_i$ adjacent to both $u$ and $v$ such that both $\widehat{v}_i u$ and $\widehat{v}_i v$ are crossing.

If $v_i$ occurs in $C_j$ positively then $u$ is on an even position and hence, $u \in V_0$. Since $\widehat{v}_i u$ is crossing it follows that $v_i$ is true and $C_j$ is satisfied. On the other hand, if $v_i$ occurs in $C_j$ negated then $u$ is on an odd position and hence, $u \in V_1$. Since $\widehat{v}_i u$ is crossing it follows that $v_i$ is false and $C_j$ is satisfied. As this holds for each clause individually, this concludes the proof. □

For every edge $e \in E(G_w)$, let $w_e$ be the weight of $e$ in $G_w$. We construct an unweighted graph $G$ from $G_w$ by replacing every edge $e = uv$ by $w_e$ paths from $u$ to $v$ on three edges. Let $W$ be the sum of the edge weights of all edges in $G_w$.

**Lemma 9.** *$G$ has a cut of size $2W + t$ if and only if $G_w$ has a cut of weight at least $t$.*

*Proof.* Given a partition of $V(G_w)$ we partition $V(G)$ as follows. The vertices of $G$ that also are vertices of $V(G)$ are partitioned in the same way as in $V(G_w)$. On each path of length 3, if the endpoints of the path are in different sets we can partition the middle vertices of the path such that all edges are cut. If the endpoints are in the same set we can only partition the middle vertices such that 2 out of the 3 edges are cut. The reverse direction is similar. □

**Lemma 10.** $\mathbf{pw}(G) \leq n + 5$.

*Proof.* We give a search strategy to clean $G$ with $n + 5$ searchers. We place one searcher on each vertex $\widehat{v}_i$ and one searcher on $x_0$. Then one can search the gadgets $\widehat{H}_j$ one by one. In $G_w$ it is sufficient to use 2 searchers for each $\widehat{H}_j$, whereas in $G$ after the edges have been replaced by multiple paths on three edges, we need 4 searchers. This combined with Proposition 1 gives the desired upper bound on the pathwidth of the graph. □

The construction, together with Lemmata 7, 8, 9 and 10 proves Theorem 3.

## 6 Graph Coloring

A $q$-coloring of $G$ is a function $\mu : V(G) \to [q]$. A $q$-coloring $\mu$ of $G$ is *proper* if for every edge $uv \in E(G)$ we have $\mu(u) \neq \mu(v)$. In the $q$-COLORING problem we are given as input a graph $G$ and the objective is to decide whether $G$ has a proper $q$-coloring. In the LIST COLORING problem, every vertex $v$ is given a list $L(v) \subseteq [q]$ of admissible colors. A *proper list coloring* of $G$ is a function $\mu : V(G) \to [q]$ such that $\mu$ is a proper coloring of $G$ that satisfies $\mu(v) \in L(v)$ for every $v \in V(G)$. In the $q$-LIST COLORING problem we are given a graph $G$ together with a list $L(v) \subseteq [q]$ for every vertex $v$. The task is to determine whether there exists a proper list coloring of $G$.

A *feedback vertex set* of a graph $G$ is a set $S \subseteq V(G)$ such that $G \setminus S$ is a forest; we denote by $\mathbf{fvs}(G)$ the size of the smallest such set. It is well-known that $\mathbf{tw}(G) \leq \mathbf{fvs}(G) + 1$. Unlike in the other sections, where we give lower bounds for algorithms parameterized by $\mathbf{pw}(G)$, the following theorem gives also a lower bound for algorithms parameterized by $\mathbf{fvs}(G)$. Such a lower bound follows very naturally from the construction we are doing here, but not from the constructions in the other sections. It would be interesting to explore whether it is possible to prove tight bounds parameterized by $\mathbf{fvs}(G)$ for the problems considered in the other sections.

**Theorem 4.** *If $q$-COLORING can be solved in $\mathcal{O}^*((q-\epsilon)^{\mathbf{fvs}(G)})$ or $\mathcal{O}^*((3-\epsilon)^{\mathbf{pw}(G)})$ time for some $\epsilon > 0$, then SAT can be solved in $\mathcal{O}^*((2-\delta)^n)$ time for some $\delta > 0$.*

**Construction.** We will show the result for LIST COLORING first, and then give a simple reduction that demonstrates that $q$-COLORING can be solved in $\mathcal{O}^*((q-\epsilon)^{\mathbf{fvs}(G)})$ time if and only if $q$-LIST COLORING can.

Depending on $\epsilon$ and $q$ we choose a parameter $p$. Now, given an instance $\phi$ to SAT we will construct a graph $G$ with a list $L(v)$ for every $v$, such that $G$ has a proper list-coloring if and only if $\phi$ is satisfiable. Throughout the construction we will call color 1-*red*, color 2-*white* and color 3-*black*.

We start by grouping the variables of $\phi$ into $t$ groups $F_1, \ldots, F_t$ of size $\lfloor \log q^p \rfloor$. Thus $t = \lceil \frac{n}{\lfloor \log q^p \rfloor} \rceil$. We will call an assignment of truth values to the variables in a group $F_i$ a *group assignment*. We will say that a group assignment satisfies a clause $C_j$ of $\phi$ if $C_j$ contains at least one literal which is set to true by the group assignment. Notice that $C_j$ can be satisfied by a group assignment of a group $F_i$, even though $C_j$ also contains variables that are not in $F_i$.

For each group $F_i$, we make a set $V_i$ of $p$ vertices $v_i^1, \ldots, v_i^p$. The vertices in $V_i$ get full lists, that is, they can be colored by any color in $[q]$. The coloring of the vertices in $V_i$ will encode the group assignment of $F_i$. There are $q^p \geq 2^{|F_i|}$ possible colorings of $V_i$. Thus, to each possible group assignment of $F_i$ we attach a unique coloring of $V_i$. Notice that some colorings of $V_i$ may not correspond to any group assignments of $F_i$.

For each clause $C_j$ of $\phi$, we make a gadget $\widehat{C}_j$. The main part of $\widehat{C}_j$ is a long path $\widehat{P}_j$ that has one vertex for each group assignment that satisfies $\widehat{C}_j$. Notice that there are at most $tq^p$ possible group assignments, and that $q$ and $p$ are constants independent of the input $\phi$. The list of every vertex on $\widehat{P}_j$ is {red, white, black}. We attach two vertices $p_j^{start}$ and $p_j^{end}$ to the start and end of $\widehat{P}_j$ respectively, and the two vertices are not counted as vertices of the path $\widehat{P}_j$ itself. The list of $p_j^{start}$ is {white}. If $|V(\widehat{P}_j)|$ is even, then the list of $p_j^{end}$ is {white}, whereas if $|V(\widehat{P}_j)|$ is odd then the list of $p_j^{end}$ is {black}. The intention is that to properly color $\widehat{P}_j$ one needs to use the color red at least once, and that once is sufficient. The position of the red colored vertex on the path $\widehat{P}_j$ encodes how the clause $C_j$ is satisfied.

For every vertex $v$ on $\widehat{P}_j$ we proceed as follows. The vertex $v$ corresponds to a group assignment to $F_i$ that satisfies the clause $C_j$. This assignment in turn corresponds to a coloring of the vertices of $V_i$. Let this coloring be $\mu_i$. We build a *connector* whose role is to enforce that $v$ can be red only if coloring $\mu_i$ appears on $V_i$. To build the connector, for each vertex $v_i^l \in V_i$ and color $x \in [q] \setminus \{\mu_i(v_i^l)\}$ we do the following.

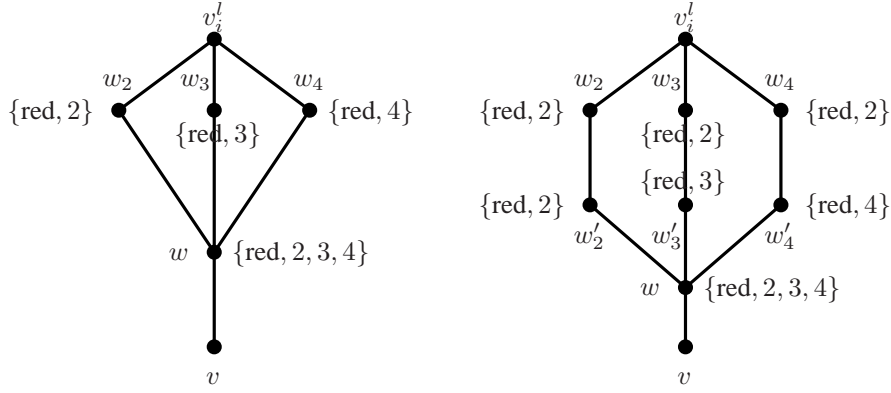

Figure 4: Reduction to $q$-COLORING: the way the connector connects a vertex $v_i^l$ with $v$ for a particular "bad color" $x \in [q] \setminus \{\mu_i(v_i^l)\}$. The left side shows the case $x = \text{red} = 1$, the right side $x = 2$ ($q = 4$).

- If $x$ is red, then we add one vertex $w_y$ for every color $y$ except for red. We make $w_y$ adjacent to $v_i^l$ and the list of $w_y$ is $\{\text{red}, y\}$. Then we add a vertex $w$ which is adjacent to all vertices $w_y$ and $v$, and whose list is all of $[q]$.

- If $x$ is not red, we add two vertices $w_y$ and $w_y'$ for each color $y$ except for red. We make $w_y$ adjacent to $v_i^l$ and $w_y'$ adjacent to $w_y$. The list of $w_y$ is $\{x, \text{red}\}$ while the list of $w_y'$ is $\{y, \text{red}\}$. Finally we add a vertex $w$ adjacent to $w_y'$ for all $y$ and to $v$. The list of $w$ is all of $[q]$.

Notice that in the above construction we have reused the names $w$, $w_y$ and $w_y'$ for many different vertices: in each connector, there is a separate vertex $w$ for each vertex $v_i^l \in V_i$ and color $x \in [q] \setminus \{\mu_i(v_i^l)\}$. Building a connector for each vertex $v$ on $\widehat{P}_j$ concludes the construction of the clause gadget $\widehat{C}_j$, and creating one such gadget for each clause concludes the construction of $G$. The following lemma summarizes the most important properties of the connector:

**Lemma 11.** *Consider the connector corresponding a vertex $v$ on $\widehat{P}_j$ and coloring $\mu_i$ of $V_i$.*

1. *Any coloring on $V_i$ and any color $c \in \{\text{white}, \text{black}\}$ on $v$ can be extended to the rest of the connector.*
2. *Coloring $\mu_i$ on $V_i$ and any color $c \in \{\text{red}, \text{white}, \text{black}\}$ on $v$ can be extended to the rest of the connector.*
3. *In any coloring of the connector, if $v$ is red, then $\mu_i$ appears on $V_i$.*

*Proof.* 1. For each vertex $v_i^l \in V_i$ and color $x \in [q] \setminus \{\mu_i(v_i^l)\}$ we do the following.

- If $x$ is red then in the construction of $\widehat{C}_j$ we added a vertex $w_y$ with list $\{y, \text{red}\}$ for every color $y \neq \text{red}$ adjacent to $v_i^l$, and a vertex $w$ with list $[q]$ adjacent to $w_y$ for every $y \neq \text{red}$. If $v_i^l$ is colored red, then we color each vertex $w_y$ with $y$ and $w$ with red. Notice that $w$ is adjacent to $v$, but $v$ is colored either white or black, so it is safe to color $w$ red. If, on the other hand, $v_i^l$ is not colored red, we can color $w_y$ red for every $y$. Then all the neighbours of $w$ have been colored with red, except for $v$ which has been colored white or black. Thus it is safe to color $w$ with the color out of black and white which was not used to color $v$.

- If $x$ is not red, then in the construction of $\widehat{C}_j$ we added two vertices $w_y$ and $w'_y$ for each color $y$ except for red, and also added a vertex $w$. The vertices $w_y$ are adjacent to $v_i^l$ and for every $y \neq$ red the vertex $w'_y$ is adjacent to $w_y$. Finally $w$ is adjacent to al the vertices $w'_y$ and to $v$. For every $y$ the list of $w_y$ is $\{x, \text{red}\}$ while the list of $w'_y$ is $\{y, \text{red}\}$. The list of $w$ is $[q]$. If $v_i^l$ is colored with $x$, then we let $w_y$ take color red and $w'_y$ take color $y$ for every $y \neq$ red. We color $w$ with red. In the case that $v_i^l$ is colored with a color different from $x$, we let $w_y$ be colored with $x$ and $w'_y$ be colored red for every $y \neq$ red. Finally, all the neighours of $w$ except for $v$ have been colored red, while $v$ is colored with either black or white. According to the color of $v$ we can either color $w$ black or white.

2. We can assume that $v$ is red, otherwise we are done by the previous statement. For each vertex $v_i^l \in V_i$ and color $x \in [q] \setminus \{\mu_i(v_i^l)\}$ we do the following.

- If $x$ is red then in the construction of $\widehat{C}_j$ we added a vertex $w_y$ with list $\{y, \text{red}\}$ for every color $y \neq$ red adjacent to $v_i^l$, and a vertex $w$ with list $[q]$ adjacent to $w_y$ for every $y \neq$ red. Since $v_{i'}^l$ is not colored red by $\mu_i$, we can color $w_y$ red for every $y$. Then all the neighbours of $w$ including $v$ have been colored with red and it is safe to color $w$ with white.

- If $x$ is not red, then in the construction of $\widehat{C}_j$ we added two vertices $w_y$ and $w'_y$ for each color $y$ except for red, and also added a vertex $w$. The vertices $w_y$ are adjacent to $v_i^l$ and for every $y \neq$ red the vertex $w'_y$ is adjacent to $w_y$. Finally $w$ is adjacent to all the vertices $w'_y$ and to $v$. For every $y$ the list of $w_y$ is $\{x, \text{red}\}$ while the list of $w'_y$ is $\{y, \text{red}\}$. The list of $w$ is $[q]$. Since $\mu_i$ colors $v_i^l$ with a color different from $x$ we let $w_y$ be colored with $x$ and $w'_y$ be colored red for every $y \neq$ red. Finally, all the neighours of $w$ including $v$ have been colored red so it is safe to color $w$ white.

3. Suppose for contradiction that $v$ is red, but some vertex $v_i^l \in V_i$ has been colored with a color $x \neq \mu_i(v_i^l)$. There are two cases. If $x$ is red, then in the construction we added vertices $w_y$ adjacent to $v_i^l$ for every color $y \neq$ red. Also we added a vertex $w$ adjacent to $v$ and to $w_y$ for each $y \neq$ red. The list of $w_y$ is $\{\text{red}, y\}$ and hence $w_y$ must have been colored $y$ for every $y \neq$ red. But then $w$ is adjacent to $v$ which is colored red, and to $w_y$ which is colored $y$ for every $y \neq$ red. Thus vertex $w$ has all colors in its neighborhood, a contradiction. In the case when $x$ is not red, then in the construction we added two vertices $w_y$ and $w'_y$ for each $y \neq$ red. Each $w_y$ was adjacent to $v_i^l$ and had $\{x, \text{red}\}$ as its list. Since $v_i^l$ is colored $x$, all the $w_y$ vertices must be colored red. For every $y \neq$ red, we have that $w'_y$ is adjacent to $w_y$ and has $\{\text{red}, y\}$ as its list. Hence for every $y \neq$ red the vertex $w'_y$ is colored with $y$. But, in the construction we also added a vertex $w$ adjacent to $v$ and to $w'_y$ for each $y \neq$ red. Thus again, vertex $w$ has all colors in its neighbourhood, a contradiction. □

**Lemma 12.** *If $\phi$ is satisfiable, then $G$ has a proper list-coloring.*

*Proof.* Starting from a satisfying assignment of $\phi$ we construct a coloring $\gamma$ of $G$. The assignment to $\phi$ corresponds to a group assignment to each group $F_i$. Each group assignment corresponds to a coloring of $V_i$. For every $i$, we let $\gamma$ color the vertices of $V_i$ using the coloring corresponding to the group assignment of $F_i$.

Now we show how to complete this coloring to a proper coloring of $G$. Since the gadgets $\widehat{C}_j$ are pairwise disjoint, and there are no edges going between them, it is sufficient to show that we can complete the coloring for every gadget $\widehat{C}_j$. Consider the clause $C_j$. The clause contains a literal that is set to true, and this literal belongs to a variable in the group $F_i$. The group assignment of $F_i$ satisfies the clause $C_j$. Thus, there is a vertex $v$ on $\widehat{P}_j$ that corresponds to this assignment. We set $\gamma(v)$ as red (that is, $\gamma$ colors $v$ red), $p_j^{start}$ is colored white and $p_j^{end}$ is colored with its only admissible color, namely black if $|V(\widehat{P}_j)|$ is even and white if $|V(\widehat{P}_j)|$ is odd. The remaining vertices of $\widehat{P}_j$ are colored alternatingly white or black. By Lemma 11(2), the coloring can be extended to every vertex of the connector between $V_i$ and $v$: the coloring appearing on $V_i$ is the coloring $\mu_i$ corresponding to the group assignment $F_i$. For every other vertex $u$ on $\widehat{P}_j$, the color of $u$ is black or white, thus Lemma 11(1) ensures that the coloring can be extended to any connector on $u$.

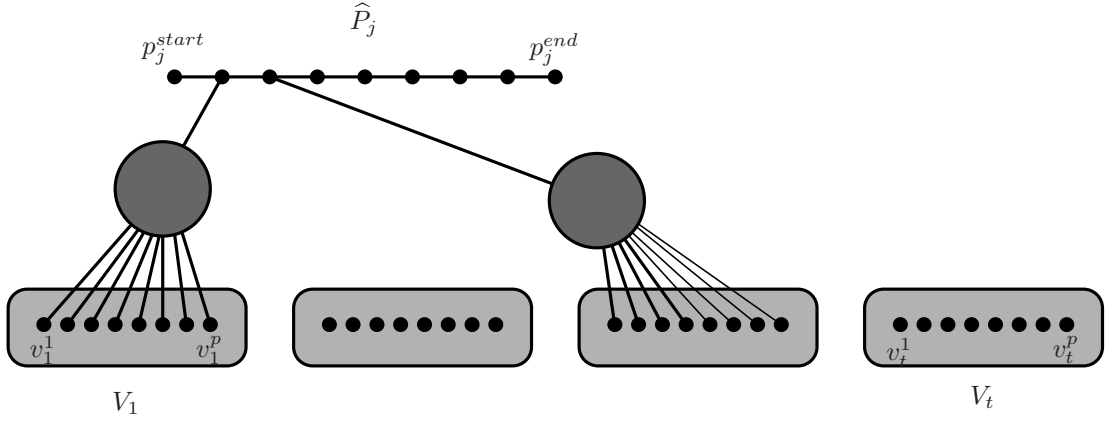

Figure 5: Reduction to $q$-COLORING. The $t$ groups of vertices $V_1, \ldots, V_t$ represent the $t$ groups of variables $F_1, \ldots, F_t$ (each of size $\lceil \log q^p \rceil$). Each vertex of the clause path $\widehat{P}_j$ is connected to one group $V_i$ via a connector.

As this procedure can be repeated to color the gadget $\widehat{C}_j$ for every clause $C_j$, we can complete $\gamma$ to a proper list-coloring of $G$. □

**Lemma 13.** *If $G$ has a proper list-coloring $\gamma$, then $\phi$ is satisfiable.*

*Proof.* Given $\gamma$ we construct an assignment to the variables of $\phi$ as follows. For every group $F_i$ of variables, if $\gamma$ colors $V_i$ with a coloring that corresponds to a group assignment of $F_i$ then we set this assignment for the variables in $F_i$. Otherwise we set all the variables in $F_i$ to false. We need to argue that this assignment satisfies all the clauses of $\phi$.

Consider a clause $C_j$ and the corresponding gadget $\widehat{C}_j$. By a simple parity argument, $\widehat{P}_j$ can not be colored using only the colors black and white. Thus, some vertex $v$ on $\widehat{P}_j$ is colored red. The vertex $v$ corresponds to a group assignment of some group $F_i$ that satisfies $\widehat{C}_j$. As $v$ is red, Lemma 11(3) implies that $V_i$ is colored with the coloring $\mu_i$ that corresponds to this assignment. The construction then implies that our chosen assignment satisfies $C_j$. As this is true for every clause, this concludes the proof. □

**Observation 1.** *The vertices $\bigcup_{i \leq t} V_i$ form a feedback vertex set of $G$. Furthermore, $\mathbf{pw}(G) \leq pt + 4$*

*Proof.* Observe that after removing $\bigcup_{i \leq t} V_i$, all that is left are the gadgets $\widehat{C}_j$ which do not have any edges between each other. Each such gadget is a tree and hence $\bigcup_{i \leq t} V_i$ form a feedback vertex set of $G$. If we place a searcher on each vertex of $\bigcup_{i \leq t} V_i$ it is easy to see that each gadget $\widehat{C}_j$ can be searched with 4 searchers. The pathwidth bound on $G$ follows using Proposition 1. □

**Lemma 14.** *If $q$-LIST COLORING can be solved in $\mathcal{O}^*((q - \epsilon)^{\mathbf{fvs}(G)})$ time for some $\epsilon < 1$, then SAT can be solved in $\mathcal{O}^*((2 - \delta)^n)$ time for some $\delta < 1$.*

*Proof.* Let $\mathcal{O}^*((q-\epsilon)^{\mathbf{fvs}(G)}) = O^*(q^{\lambda \mathbf{fvs}(G)})$ time, where $\lambda = \log_q(q-\epsilon) < 1$. We choose a sufficiently large $p$ such that $\delta' = \lambda \frac{p}{p-1} < 1$. Given an instance $\phi$ of SAT we construct a graph $G$ using the construction above, and run the assumed $q$-LIST COLORING. Correctness follows from Lemmata 12 and 13. By Observation 1 the graph $G$ has a feedback vertex set of size $p \lceil \frac{n}{\lfloor p \log q \rfloor} \rceil$. The choice of $p$ implies that

$$\lambda p \lceil \frac{n}{\lfloor p \log q \rfloor} \rceil \leq \lambda p \frac{n}{(p-1) \log q} + p \leq \delta' \frac{n}{\log q} + p \leq \delta'' n,$$

for some $\delta'' < 1$. Hence SAT can be solved in time $\mathcal{O}^*(2^{\delta''n}) = \mathcal{O}^*((2-\delta)^n)$, for some $\delta > 0$. □

Finally, observe that we can reduce $q$-LIST-COLORING to $q$-COLORING by adding a clique $Q = \{q_1, \ldots, q_c\}$ on $q$ vertices to $G$ and making $q_i$ adjacent to $v$ when $i \notin L(v)$. Any coloring of $Q$ must use $q$ different colors, and without loss of generality $q_i$ is colored with color $i$. Then one can complete the coloring if and only if one can properly color $G$ using a color from $L(v)$ for each $v$. We can add the clique $Q$ to the feedback vertex set—this increases the size of the minimum feedback vertex set by $q$. Since $q$ is a constant independent of the input, this yields Theorem 4.

## 7 Odd Cycle Transversal

An equivalent formulation of MAX CUT is to delete the minimum number of edges to make the graph bipartite. We can also consider the vertex deletion version of the problem. An *odd cycle transversal* of a graph $G$ is a subset $S \subseteq V(G)$ such that $G \setminus S$ is bipartite. In the ODD CYCLE TRANSVERSAL problem we are given a graph $G$ together with an integer $k$ and asked whether $G$ has an odd cycle transversal of size $k$.

**Theorem 5.** *If* ODD CYCLE TRANSVERSAL *can be solved in* $\mathcal{O}^*((3-\epsilon)^{\mathbf{pw}(G)})$ *time for* $\epsilon > 0$, *then* SAT *can be solved in* $\mathcal{O}^*((2-\delta)^n)$ *time for some* $\delta > 0$.

**Construction.** Given $\epsilon > 0$ and an instance $\phi$ of SAT we construct a graph $G$ as follows. We chose an integer $p$ based just on $\epsilon$. Exactly how $p$ is chosen will be discussed at the end of this section. We start by grouping the variables of $\phi$ into $t$ groups $F_1, \ldots, F_t$ of size at most $\lfloor \log 3^p \rfloor$. Thus $t = \lceil \frac{n}{\lfloor \log 3^p \rfloor} \rceil$. We will call an assignment of truth values to the variables in a group $F_i$ a *group assignment*. We will say that a group assignment satisfies a clause $C_j$ of $\phi$ if $C_j$ contains at least one literal which is set to true by the group assignment. Notice that $C_j$ can be satisfied by a group assignment of a group $F_i$, even though $C_j$ also contains variables that are not in $F_i$.

Now we describe an auxiliary gadget which will be very useful in our construction. For two vertices $u$ and $v$ by *adding an arrow* from $u$ to $v$ we will mean adding a path $ua_1a_2a_3v$ on four edges starting in $u$ and ending in $v$. Furthermore, we add four vertices $b_1$, $b_2$, $b_3$ and $b_4$ and edges $ub_1$, $b_1a_1$, $a_1b_2$, $b_2a_2$, $a_2b_3$, $b_3a_3$, $a_3b_4$, $b_4v$, and $b_4v$. Denote the resulting graph $A(u,v)$. None of the vertices in $A(u,v)$ except for $u$ and $v$ will receive any further neighbours throughout the construction of $G$. The graph $A(u,v)$ has the following properties, which are useful for our construction.

- The unique smallest odd cycle transversal of $A(u,v)$ is $\{a_1, a_3\}$. We call this the *passive* odd cycle transversal of the arrow.
- In $A(u,v) \setminus \{a_1, a_3\}$, $u$ and $v$ are in different connected components.
- The set $\{a_2, v\}$ is a smallest odd cycle transversal of $A(u,v) \setminus \{u\}$. We call this the *active* odd cycle transversal of the arrow.

The intuition behind an arrow from $u$ to $v$ is that if $u$ is put into the odd cycle transversal, then $v$ can be put into the odd cycle transversal "for free." When the active odd cycle transversal of the arrow is picked, we say the arrow is active, otherwise we say the arrow is passive.

To construct $G$ we make $t \cdot p$ paths, $\{P_{i,j}\}$ for $1 \le i \le t$, $1 \le j \le p$. Each path has $3m(tp+1)$ vertices, and the vertices of $P_{i,j}$ are denoted by $p_{i,j}^\ell$ for $1 \le \ell \le 3m(tp+1)$. For a fixed $i$, the paths $\{P_{i,j} : 1 \le j \le p\}$ correspond to the set $F_i$ of variables. For every $1 \le i \le t$, $1 \le j \le p$ and $1 \le \ell < 3m(tp+1)$ we add three vertices $a_{i,j}^\ell$, $b_{i,j}^\ell$ and $q_{i,j}^\ell$ adjacent to each other. We also add the edges $a_{i,j}^\ell p_{i,j}^\ell$ and $b_{i,j}^\ell p_{i,j}^{\ell+1}$.

One can think of the vertices of the paths $\{P_{i,j}\}$ layed out as rows in a matrix, where for every fixed $1 \le \ell \le 3m(tp+1)$ there is a column $\{p_{i,j}^\ell : 1 \le i \le t, 1 \le j \le p\}$. We group the colums three by three.

In particular, For every $i \leq t$ and $0 \leq \ell < m(tp+1)$ we define the sets $P_i^\ell = \{p_{i,j}^{3\ell+1}, p_{i,j}^{3\ell+2}, p_{i,j}^{3\ell+3} : 1 \leq j \leq p\}$, $A_i^\ell = \{a_{i,j}^{3\ell+1}, a_{i,j}^{3\ell+2}, a_{i,j}^{3\ell+3} : 1 \leq j \leq p\}$, $B_i^\ell = \{b_{i,j}^{3\ell+1}, b_{i,j}^{3\ell+2}, b_{i,j}^{3\ell+3} : 1 \leq j \leq p\}$ and $Q_i^\ell = \{q_{i,j}^{3\ell+1}, q_{i,j}^{3\ell+2}, q_{i,j}^{3\ell+3} : 1 \leq j \leq p\}$.

For every $i \leq t$ and $0 \leq \ell < m(tp+1)$ we make two new sets $L_i^\ell$ and $R_i^\ell$ of new vertices. Both $L_i^\ell$ and $R_i^\ell$ are independent sets of size $5p$, and we add all the edges possible between $L_i^\ell$ and $R_i^\ell$. From $L_i^\ell$ we pick a special vertex $\widehat{l}_i^\ell$ and from $R_i^\ell$ we pick $\widehat{r}_i^\ell$. We make all the vertices in $A_i^\ell$ adjacent to all vertices of $L_i^\ell$, and we make all vertices in $B_i^\ell$ adjacent to all vertices of $R_i^\ell$. We make $l_i^\ell$ adjacent to $r_i^{\ell+1}$, except for $\ell = m(tp+1) - 1$.

We will say that a subset $S$ of $P_i^\ell$ which picks exactly one vertex from $P_{i,j}$ for every $1 \leq j \leq p$ is *good*. The idea is that there are $3^p \geq 2^h$ good subsets of $P_i^\ell$, so we can make group assignments of $F_i$ correspond to good subsets of $P_i^\ell$. For every good subset $S$ of $P_i^\ell$ we add a cycle $X_{i,S}^\ell$. The cycle $X_{i,S}^\ell$ has length $2p+1$. We select a vertex on $X_{i,S}^\ell$ and call it $x_{i,S}^\ell$. For every vertex $u \in P_i^\ell \setminus S$ we add an arrow from $u$ to a vertex of $X_{i,S}^\ell$. We add arrows in such a way that every vertex of $X_{i,S}^\ell$ is the endpoint of exactly one arrow.

For every $i \leq t$ and $0 \leq \ell < m(tp+1)$ we make a cycle $Y_i^\ell$ of length $3^p$, notice that the length of the cycle is odd. Every vertex of $Y_i^\ell$ corresponds to a good subset $S$ of $P_i^\ell$. For each good subset $S$ of $P_i^\ell$ we add an arrow from $x_{i,S}^\ell$ of the cycle $X_{i,S}^\ell$ to the vertex in $Y_i^\ell$ which corresponds to $S$.

We say that a good subset of $P_i^\ell$ is *equal* with a good subset $S'$ of $P_i^{\ell'}$ if for every $1 \leq j \leq t$, the distance along $P_{i,j}$ between the vertex of $S$ on $P_{i,j}$ and the vertex of $S'$ on $P_{i,j}$ is divisible by $3$. Informally, $S$ and $S'$ are equal if they look identical when we superimpose $P_i^\ell$ onto $P_i^{\ell'}$. To every group assignment of variables $F_i$ we designate a good subset of $P_i^\ell$ for every $\ell$. We designate good subsets in such a way that good subsets corresponding to the same group assignment are equal.

Finally, for every clause $C_h$, $1 \leq h \leq m$, we will add $tp+1$ cycles. That is, for every $0 \leq r \leq tp$ we add a cycle $\widehat{C}_j^r$. The cycle contains one vertex for every $i \leq t$ and group assignment to $F_i$, and potentially one dummy vertex to make it have odd length. Going around the cycle counterclockwise we first encounter all the vertices corresponding to group assignments of $F_1$, then all the vertices corresponding to group assignments of $F_2$, and so on. For $i \leq t$ and every good subset $S$ of $P_i^{rm+j}$ that corresponds to a group assignment of $F_i$ that satisfies $C_j$ we add an arrow from $x_{i,S}^{rm+j}$ to the vertex on $\widehat{C}_j^r$ that corresponds to the same group assignment of $F_i$ as $S$ does. This concludes the construction of $G$.

The intention behind the construction is that if $\phi$ is satisfiable, then a minimum odd cycle transversal of $G$ can pick:

- One vertex from each triangle $\{a_{i,j}^\ell, b_{i,j}^\ell, q_{i,j}^\ell\}$ for each $1 \leq i \leq t, 1 \leq j \leq p, 1 \leq \ell < 3m(tp+1)$. There are $tp(3m(tp+1)-1)$ such triangles in total.
- One vertex from $\{p_{i,j}^{3\ell+1}, p_{i,j}^{3\ell+2}, p_{i,j}^{3\ell+3}\}$ for each $1 \leq i \leq t, 1 \leq j \leq p, 0 \leq \ell < m(tp+1)$. There are $tpm(tp+1)$ such triples.
- Two vertices from every arrow added, *without* counting the starting point of the arrow. For each $i \leq t$ and $0 \leq \ell < m(tp+1)$ there are $2p3^p$ arrows ending in some cycle $X_{i,S}^\ell$. Hence there are $2p3^p tm(tp+1)$ such arrows. For every $i \leq t$ and $0 \leq \ell < m(tp+1)$ there are $3^p$ arrows ending in the cycle $Y_i^\ell$. Hence there are $3^p tm(tp+1)$ such arrows. For every clause $C_j$ there are $m$ arrows added for every group assignment that satisfies that clause. Let $\mu$ be the sum over all clauses of the number of group assignments that satisfy that clause. The total number of arrows added is then $m\mu + (2p+1)3^p tm(tp+1)$. Thus the odd cycle transversal can pick $2m\mu + 2(2p+1)3^p tm(tp+1)$ vertices from arrows.
- One vertex $x_{i,S}^\ell$ for every $i \leq t$ and $0 \leq \ell < m(tp+1)$. There are $tm(tp+1)$ choices for $i$ and $\ell$.

We let the $\alpha$ be the value of the total budget, that is the sum of the items above.

**Lemma 15.** *If $\phi$ is satisfiable, then $G$ has an odd cycle transversal of size $\alpha$.*

*Proof.* Given a satisfying assignment $\gamma$ to $\phi$ we construct an odd cycle transversal $Z$ of $G$ of size $\alpha$ together with a partition of $V(G) \setminus Z$ into $L$ and $R$ such that every edge of $G \setminus Z$ goes between a vertex in $L$ and a vertex in $R$. The assignment to $\phi$ corresponds to a group assignment of each $F_i$ for $1 \leq i \leq t$. For every $1 \leq i \leq t$ and $0 \leq \ell < m(tp+1)$ we add to $Z$ the good subset $S$ of $P_i^\ell$ that corresponds to the group assignment of $F_i$. Notice that for each fixed $i$, the sets picked from $P_i^\ell$ and $P_i^{\ell'}$ are equal for any $\ell, \ell'$. At this point we have picked one vertex from $\{p_{i,j}^{3\ell+1}, p_{i,j}^{3\ell+2}, p_{i,j}^{3\ell+3}\}$ for each $1 \leq i \leq t$, $1 \leq j \leq p$, $0 \leq \ell < m(tp+1)$.

For every fixed $1 \leq i \leq t$, $1 \leq j \leq p$ there are three cases. If $p_{i,j}^1 \in Z$ we put $p_{i,j}^2$ into $L$ and $p_{i,j}^3$ into $R$. If $p_{i,j}^2 \in Z$ we put $p_{i,j}^1$ into $R$ and $p_{i,j}^3$ into $L$. If $p_{i,j}^3 \in Z$ we put $p_{i,j}^1$ into $L$ and $p_{i,j}^2$ into $R$. Now, for every $4 \leq \ell \leq 3m(tp+1)$ such that $p_{i,j}^\ell \notin Z$ we put $p_{i,j}^\ell$ into the same set out of $\{L, R\}$ as $p_{i,j}^{\ell'}$ where $1 \leq \ell' \leq 3$ and $\ell \equiv \ell' \mod 3$.

For every $1 \leq i \leq t$, $0 \leq \ell \leq m(tp+1)$ we put $L_i^\ell$ into $L$ and $R_i^\ell$ into $R$. For every triple of $a, b, q$ of pairwise adjacent vertices such that $a \in A_i^\ell$, $b \in B_i^\ell$, and $q \in Q_i^\ell$, we proceed as follows. The vertex $a$ has a neighbour $a'$ in $P_i^\ell$ and $b$ has a neighbour $b'$ in $P_i^\ell$. There is a $j$ such that $b'$ is the successor of $a'$ on $P_{i,j}$. Thus, there are three cases;

- $a' \in Z$ and $b' \in L$, we put $a$ in $R$, $q$ in $L$ and $b$ in $Z$.
- $a' \in R$ and $b' \in Z$, we put $a$ in $Z$, $q$ in $R$ and $b$ in $L$.
- $a' \in L$ and $b' \in R$, we put $a$ in $R$, $q$ in $Z$ and $b$ in $L$.

For every $1 \leq i \leq t$, $0 \leq \ell \leq m(tp+1)$ there are many arrows from vertices in $P_i^\ell$ to vertices on cycles $X_{i,S}^\ell$ for good subsets $S$ of $P_i^\ell$. For each arrow, if its endpoint in $P_i^\ell$ is in $Z$ we add the active odd cycle transversal of the arrow to $Z$, otherwise we add the passive odd cycle transversal of the arrow to $Z$. In either case the remaining vertices on the arrow form a forest, and therefore we can insert the remaining vertices of the arrow into $L$ and $R$ according to which sets out of $\{L, R, Z\}$ $u$ and $v$ are in.

For every $1 \leq i \leq t$, $0 \leq \ell \leq m(tp+1)$ there is exactly one set $S$ such that the cycle $X_{i,S}^\ell$ only has passive arrows pointing into it. This is exactly the set $S$ which corresponds to the restriction of $\gamma$ to $F_i$. Each cycle $X_{i,S'}^\ell$ that has at least one arrow pointing into them already contain at least one vertex in $Z$—the endpoint of the active arrow pointing into the cycle. Thus we can partition the remaining vertices of $X_{i,S'}^\ell$ into $L$ and $R$ such that no edge has both endpoints in $L$ or both endpoints in $R$. For the cycle $X_{i,S}^\ell$ we put $x_{i,S}^\ell$ into $Z$ and partition the remaining vertices of $X_{i,S}^\ell$ into $L$ and $R$ such that no edge has both endpoints in $L$ or both endpoints in $R$. We add the active odd cycle transversal in the arrow from $x_{i,S}^\ell$ to the cycle $Y_i^\ell$ into $Z$. For all other good subsets $S'$ we add the passive odd cycle transversal in the arrow from $x_{i,S}^\ell$ to the cycle $Y_i^\ell$ into $Z$. Thus each cycle $Y_i^\ell$ contains one vertex in $Z$ and the remaining vertices of $Y_i^\ell$ can be distributed into $L$ and $R$.

For every arrow that goes from a vertex $x_{i,S}^\ell$ into a cycle $\widehat{C}_h^r$ we add the active odd cycle transversal of the arrow to $Z$ if $x_{i,S}^\ell \in Z$ and add the passive odd cycle transversal to $Z$ otherwise. Again the remaining vertices on each arrow can easily be partitioned into $L$ and $R$ such that no edge has both endpoints in $L$ or both endpoints in $R$. This concludes the construction of $Z$. Since we have put the vertices into $Z$ in accordance to the budget described in the construction it follows that $|Z| \leq \alpha$. All that remains to show, is that for each $1 \leq h \leq m$ and $0 \leq r < n+1$, the cycle $\widehat{C}_h^r$ has at least one active arrow pointing into it.

The cycle $\widehat{C}_h^r$ corresponds to the clause $C_h$. The clause $C_h$ is satisfied by $\gamma$ and hence it is satisfied by the restriction of $\gamma$ to a group $F_i$. This restriction is a group assignment of $F_i$ and hence it corresponds to a good subset $S$ of $P_i^{rm+h}$, which happens to be exactly $Z \cap P_i^{rm+h}$. Thus $x_{i,S}^{rm+h} \in Z$ and since the restriction of $\gamma$ to $F_i$ satisfies $C_h$ there is an arrow pointing from $x_{i,S}^{rm+h}$ and into $\widehat{C}_h^r$. Since this arrow is active, this concludes the proof. □

**Lemma 16.** *If $G$ has an odd cycle transversal of size $\alpha$, then $\phi$ is satisfiable.*

*Proof.* Let $Z$ be an odd cycle transversal of $G$ of size $\alpha$. Since $G \setminus Z$ is bipartite, the vertices of $G \setminus Z$ can be partitioned into $L$ and $R$ such that every edge of $G \setminus Z$ has one endpoint in $L$ and the other in $R$. Given $Z$, $L$ and $R$, we construct a satisfying assignment to $\phi$. Every arrow in $G$ must contain at least two vertices in $Z$, not counting the startpoint of the arrow. Let $\vec{Z}$ be a subset of $Z$ containing two vertices from each arrow, but no arrow start point. Observe that no two arrows have the same endpoint, and therefore $|\vec{Z}|$ is exactly two times the number of arrows in $G$. Let $Z' = Z \setminus \vec{Z}$.

We argue that for any $1 \leq i \leq t$ and $0 \leq \ell < m(tp+1)$ we have $|Z' \cap (L_i^\ell \cup R_i^\ell \cup A_i^\ell \cup B_i^\ell \cup Q_i^\ell \cup P_i^\ell)| \geq 4p$. Observe that no vertices in $L_i^\ell, R_i^\ell, A_i^\ell, B_i^\ell, Q_i^\ell$ or $P_i^\ell$ are endpoints of arrows, and hence they do not contain any vertices of $\vec{Z}$. Suppose for contradiction that $|Z' \cap (L_i^\ell \cup R_i^\ell \cup A_i^\ell \cup B_i^\ell \cup Q_i^\ell \cup P_i^\ell)| < 4p$. Then there is a vertex in $\widehat{l} \in L_i^\ell \setminus Z'$, and a vertex $\widehat{r} \in R_i^\ell \setminus Z'$. Without loss of generality, $\widehat{l} \in L$ and $\widehat{r} \in R$. Furthermore, there is a $1 \leq j \leq p$ such that

$$|Z' \cap \{p_{i,j}^{3\ell+1}, p_{i,j}^{3\ell+2}, p_{i,j}^{3\ell+3}, a_{i,j}^{3\ell+1}, a_{i,j}^{3\ell+2}, a_{i,j}^{3\ell+3}, b_{i,j}^{3\ell+1}, b_{i,j}^{3\ell+2}, b_{i,j}^{3\ell+3}, q_{i,j}^{3\ell+1}, q_{i,j}^{3\ell+2}, q_{i,j}^{3\ell+3}\}| < 4.$$

Since $\{a_{i,j}^{3\ell+1}, b_{i,j}^{3\ell+1}, c_{i,j}^{3\ell+1}\}$, $\{a_{i,j}^{3\ell+2}, b_{i,j}^{3\ell+2}, c_{i,j}^{3\ell+2}\}$ and $\{a_{i,j}^{3\ell+3}, b_{i,j}^{3\ell+3}, c_{i,j}^{3\ell+3}\}$ form triangles and must contain a vertex from $Z'$ each, it follows that each of these triangles contain exactly one vertex from $Z'$, and that $Z' \cap \{p_{i,j}^{3\ell+1}, p_{i,j}^{3\ell+2}, p_{i,j}^{3\ell+3}\} = \emptyset$. Since $\widehat{l} \in L$ and $\widehat{r} \in R$, $\widehat{l}$ is adjacent to all vertices of $A_{i,j}^\ell$ and $\widehat{r}$ is adjacent to all vertices of $B_{i,j}^\ell$ it follows that $A_{i,j}^\ell \setminus Z' \subseteq R$ and $B_{i,j}^\ell \setminus Z' \subseteq L$.

Hence, there are two cases to consider either (1) $\{p_{i,j}^{3\ell+1}, p_{i,j}^{3\ell+3}\} \subseteq L$ and $p_{i,j}^{3\ell+2} \in R$ or (2) $\{p_{i,j}^{3\ell+1}, p_{i,j}^{3\ell+3}\} \subseteq R$ and $p_{i,j}^{3\ell+2} \in L$. In the first case observe that either $a_{i,j}^{3\ell+2} \in R$ or $b_{i,j}^{3\ell+2} \in L$ and hence either $a_{i,j}^{3\ell+2} p_{i,j}^{3\ell+2}$ or $b_{i,j}^{3\ell+2} p_{i,j}^{3\ell+3}$ have both endpoints in the same set out of $\{L, R\}$, a contradiction. The second case is similar, either $a_{i,j}^{3\ell+1} \in R$ or $b_{i,j}^{3\ell+1} \in L$ and hence either $a_{i,j}^{3\ell+1} p_{i,j}^{3\ell+1}$ or $b_{i,j}^{3\ell+1} p_{i,j}^{3\ell+2}$ have both endpoints in the same set out of $\{L, R\}$, a contradiction. We conclude that $|Z' \cap (L_i^\ell \cup R_i^\ell \cup A_i^\ell \cup B_i^\ell \cup Q_i^\ell \cup P_i^\ell)| \geq 4p$.

For any $1 \leq i \leq t$ and $0 \leq \ell < m(tp+1)$, $Y_i^\ell$ is an odd cycle so $Y_i^\ell$ contains a vertex in $Z$. If $Y_i^\ell$ contains no vertices of $Z'$ it contains a vertex from $\vec{Z}$ and there is an active arrow pointing into $Y_i^\ell$. The starting point of this arrow is a vertex $x_{i,S}^\ell$ for some good subset $S$ of $P_i^\ell$. Since the arrow is active and $x_{i,S}^\ell$ is not the endpoint of any arrow, we know that $x_{i,S}^\ell \in Z'$. Hence for any $1 \leq i \leq t$ and $0 \leq \ell < m(tp+1)$ we have that either there is a good subset $S$ of $P_i^\ell$ such that $x_{i,S}^\ell \in Z'$ or at least one vertex of $Y_i^\ell$ is in $Z'$.

The above arguments, together with the budget constraints, imply that for every $1 \leq i \leq t$ and $0 \leq \ell < m(tp+1)$, we have $|Z' \cap (L_i^\ell \cup R_i^\ell \cup A_i^\ell \cup B_i^\ell \cup Q_i^\ell \cup P_i^\ell)| = 4p$ and that $|Z' \cap \bigcup \{x_{i,S}^\ell\} \cup V(Y_i^\ell)| = 1$, where the union is taken over all good subsets $S$ of $P_i^\ell$. It follows $Z' \cap P_i^\ell$ is a good subset of $P_i^\ell$. Let $S = Z' \cap P_i^\ell$. The cycle $X_{i,S}^\ell$ has odd length, and hence it must contain some vertex from $Z$. On the other hand, all the arrows pointing into $X_{i,S}^\ell$ are passive, so $X_{i,S}^\ell$ cannot contain any vertices from $\vec{Z}$. Thus $X_{i,S}^\ell$ contains a vertex from $Z'$, and by the budget constraints this must be $x_{i,S}^\ell$.

Now, consider three consecutive vertices $p_{i,j}^\ell, p_{i,j}^{\ell+1}, p_{i,j}^{\ell+2}$ for some $1 \leq i \leq t$, $1 \leq j \leq p$, $1 \leq \ell \leq 3m(tp+1) - 2$. We prove that at least one of them has to be in $Z$. Suppose not. We know that neither $\widehat{l}_i^{\lfloor \ell/3 \rfloor}, \widehat{r}_i^{\lfloor \ell/3 \rfloor}, \widehat{l}_i^{\lfloor \ell/3 \rfloor+1}$ nor $\widehat{r}_i^{\lfloor \ell/3 \rfloor+1}$ are in $Z$. Thus, without loss of generality $\{\widehat{l}_i^{\lfloor \ell/3 \rfloor}, \widehat{l}_i^{\lfloor \ell/3 \rfloor+1}\} \subseteq L$ and $\{\widehat{r}_i^{\lfloor \ell/3 \rfloor}, \widehat{r}_i^{\lfloor \ell/3 \rfloor+1}\} \subseteq R$. There are two cases. Either $p_{i,j}^\ell \in R$ and $p_{i,j}^{\ell+1} \in L$ or $p_{i,j}^{\ell+1} \in L$ and $p_{i,j}^{\ell+3} \in R$. In the first case we obtain a contradiction since either $a_{i,j}^\ell \in R$ or $b_{i,j}^\ell \in L$. In the second case we get a contradiction since either $a_{i,j}^{\ell+1} \in R$ or $b_{i,j}^{\ell+1} \in L$. Hence for any three consecutive vertices on $P_{i,j}$, at least one of them is in $Z$. Since the budget constraints ensure that there are at most $|V(P_{i,j})|/3$ vertices in $P_{i,j} \cap Z$ it follows from the pigeon hole principle, that there is an $0 \leq r < n+1$ such that for any $1 \leq i \leq t$ and $1 \leq h \leq m$ and $1 \leq h' \leq m$ the set $P_i^{rm+h} \cap Z$ equals $P_i^{rm+h'} \cap Z$. Here equality is in the sense of equality of good subsets of $P_i^\ell$.

For every $1 \leq i \leq t$, $P_i^{rm+1} \cap Z$ is a good subset of $P_i^{rm+1}$. If $P_i^{rm+1} \cap Z$ corresponds to a group assignment of $F_i$, then we set the variables in $F_i$ to this assignment. Otherwise we set all the variables in $F_i$ to false. We need to argue that every clause $C_h$ is satisfied by this assignment. Consider the cycle $\widehat{C}_h^r$. Since it is an odd cycle, it must contain a vertex from $Z$, the budget constraints and the discussion above implies that this vertex is from $\vec{Z}$. Hence there must be an active arrow pointing into $\widehat{C}_h^r$. The starting point of this active arrow is a vertex $x_{i,S}^{mr+h}$ for some $i$ and good subset $S$ of $P_i^{mr+h}$. The set $S$ corresponds to a group assignment of $F_i$ that satisfies $C_h$. Since the arrow is active $x_{i,S}^{mr+h} \in Z'$, and by the discussion above we have that $P_i^{mr+h} \cap Z' = S$. Now, $S = P_i^{mr+h} \cap Z'$ and $S$ is equal to $P_i^{mr+1} \cap Z'$ and hence the assignment to the variables of $F_i$ satisfies $C_h$. Since this holds for all clauses, this concludes the proof. □

**Lemma 17.** $\mathbf{pw}(G) \leq t(p+1) + 10p3^p$.

*Proof.* We show how to search the graph using at most $t(p+1) + 10p3^p$ searchers. The strategy consists of $m(tp+1)$ rounds numbered from round 0 to round $m(tp+1)-1$. Each round has $t$ stages, numbered from 1 to $t$. In the beginning of round $k$ there is a searcher on $p_{i,j}^{3k+1}$ and $\widehat{r}_i^k$ for every $1 \leq i \leq t$, $1 \leq j \leq p$. Let $r$ and $1 \leq h \leq m$ be integers such that $k+1 = rm + h$. Recall, that as we go around $\widehat{C}_h^r$ counterclockwise we first encounter vertices corresponding to group assignments of $F_1$, then to assignments of $F_2$ and so on. In the beginning of round $k$ we place a searcher on the first vertex on $\widehat{C}_h^r$ that corresponds to an assignment of $F_1$. If $\widehat{C}_h^r$ contains a dummy vertex, we place a searcher on this vertex as well. These two searchers will remain on their respective vertices throughout the round. In the beginning of stage $s$ of round $k$ we will assume that the vertices on the cycle $\widehat{C}_h^r$ corresponding to group assignments of $F_{s'}$, $s' < s$ have already been cleaned, and in the beginning of every stage $s > 1$, there is a searcher standing on the first vertex corresponding to a group assignment of $F_s$.

In stage $s$ of round $k$, we place searchers on all vertices of $P_s^k, A_s^k, B_s^k, Q_s^k, L_s^k, R_s^k, Y_s^k$ and all vertices of cycles $X_{s,S}^k$ for every good subset $S$ of $P_s^k$, on all vertices of arrows starting or ending in such cycles, and on all vertices of $\widehat{C}_h^r$ corresponding to group assignments of $F_s$. In total this amounts to less than $10p3^p$ vertices.

In the last part of stage $s$ of round $k$, we place searchers on $p_{s,j}^{3(k+1)+1}$ for every $1 \leq j \leq p$ and on $\widehat{r}_s^{k+1}$. Then we remove all the searchers that were placed out in the first part of phase $s$ except for the searcher on the last vertex on $\widehat{C}_h^r$ corresponding to a group assignment of $F_s$. Unless $s = 1$ there is also a searcher on the last vertex on $\widehat{C}_h^r$ corresponding to a group assignment of $F_{s-1}$. We remove this searcher, and the next stage can commence. In the end of the last stage of round $k$ we remove all the searchers from $\widehat{C}_h^r$. Then the last stage can commence. At any point in time, at most $t(p+1) + 10p3^p$ searchers are placed on $G$. □

*Proof (of Theorem 5).* Suppose ODD CYCLE TRANSVERSAL can be solved in $\mathcal{O}^*((3-\epsilon)^{\mathbf{pw}(G)})$ time for $\epsilon < 1$. Then there is an $\epsilon' < 1$ such that $\mathcal{O}^*((3-\epsilon)^{\mathbf{pw}(G)}) \leq \mathcal{O}^*(3^{\epsilon' \mathbf{pw}(G)})$. We chose $p$ large enough such that $\epsilon' \cdot \frac{p+1}{p-1} = \delta' < 1$. Given an instance of SAT we construct an instance of ODD CYCLE TRANSVERSAL using the above construction and the chosen value of $p$. Then we solve the ODD CYCLE TRANSVERSAL instance using the $\mathcal{O}^*((3-\epsilon)^{\mathbf{pw}(G)})$ time algorithm. Correctness is ensured by Lemmata 15 and 16. Lemma 17 yields that the total time taken is upper bounded by $\mathcal{O}^*((3-\epsilon)^{\mathbf{pw}(G)}) \leq \mathcal{O}^*(3^{\epsilon' \mathbf{pw}(G)}) \leq \mathcal{O}^*(3^{\epsilon'(t(p+1)+f(\epsilon'))}) \leq \mathcal{O}^*(3^{\epsilon' \lceil \frac{n}{p \log 3} \rceil (p+1)}) \leq \mathcal{O}^*(3^{\epsilon' \frac{n(p+1)}{\lceil p \log 3 \rceil}}) \leq \mathcal{O}^*(3^{\epsilon' \frac{n(p+1)}{(p-1) \log 3}}) \leq \mathcal{O}^*(3^{\delta' \frac{n}{\log 3}}) \leq \mathcal{O}^*(2^{\delta' n}) =. \mathcal{O}^*((2-\delta)^n)$ for $\delta < 1$. □

## 8 Partition Into Triangles

A *triangle packing* in a graph $G$ is a collection of pairwise disjoint vertex sets $S_1, S_2, \ldots S_t$ in $G$ such that $S_i$ induces a triangle in $G$ for every $i$. The size of the packing is $t$. If $V(G) = \bigcup_{i \leq t} S_i$ then the collection $S_1 \ldots S_t$ is a *partition of $G$ into triangles*. In the TRIANGLE PACKING problem we are given a graph $G$ and an integer $t$ and asked whether there is a triangle packing in $G$ of size at least $t$. In the PARTITION INTO TRIANGLES problem we are given a graph $G$ and asked whether $G$ can be partitioned into triangles. Notice that since PARTITION INTO TRIANGLES is the special case of TRIANGLE PACKING when the number of triangles is the number of vertices divided by 3, the bound of Theorem 6 holds for TRIANGLE PACKING as well.

**Theorem 6.** *If* PARTITION INTO TRIANGLES *can be solved in* $\mathcal{O}^*((2-\epsilon)^{\mathbf{pw}(G)})$ *for* $\epsilon > 0$ *then* SAT *can be solved in* $\mathcal{O}^*((2-\delta)^n)$ *time for some* $\delta > 0$.

**Construction.** first show the lower bound for TRIANGLE PACKING and then modify our construction to also work for the more restricted PARTITION INTO TRIANGLES problem. Given an instance $\phi$ of SAT we construct a graph $G$ as follows. For every variable $v_i$ we make a path $P_i$ on $2m(n+1)+1$ vertices. We denote the $l$'th vertex of $P_i$ by $p_i^l$. For every $i$ we add a set $T_i$ of $2m(n+1)$ vertices, and let the $l$'th vertex of $T_i$ be denoted $t_i^l$. For every $1 \le l \le 2m(n+1)$ we add the edges $t_i^l p_i^l$ and $t_i^l p_i^{l+1}$.

For every clause $C_j$ we add $n+1$ gadgets corresponding to the clause. In particular, for every $0 \le r \le n$ we do the following. First we add the vertices $\widehat{c}_j^r$ and $\widehat{d}_j^r$ and the edge $\widehat{c}_j^r \widehat{d}_j^r$. For every variable $v_i$ that occurs in $C_j$ positively we add the edges $\widehat{c}_j^r t_i^{2(mr+j)}$ and $\widehat{d}_j^r t_i^{2(mr+j)}$. For every variable $v_i$ that occurs in $C_j$ negated we add the edges $\widehat{c}_j^r t_i^{2(mr+j)-1}$ and $\widehat{d}_j^r t_i^{2(mr+j)-1}$. Doing this for every $r$ and every clause $C_j$ concludes the construction of $G$.

**Lemma 18.** *If $\phi$ satisfiable, then $G$ has a triangle packing of size $mn(n+1) + m(n+1)$.*

*Proof.* Consider a satisfying assignment to $\phi$. For every variable $v_i$ that is set to true and integer $1 \le l \le m(n+1)$ we add $\{t_i^{2l-1}, p_i^{2l-1}, p_i^{2l}\}$ to the triangle packing. For every variable $v_i$ that is set to false and integer $1 \le l \le m(n+1)$ we add $\{t_i^{2l}, p_i^{2l}, p_i^{2l+1}\}$ to the triangle packing. For every clause $C_j$ there is a literal set to true. Suppose this literal corresponds to the variable $v_i$. Notice that if $v_i$ occurs positively in $C_j$, then $v_i$ is set to true, and if it occurs negatively it is set to false. For each $0 \le r \le n$, if $v_i$ occurs positively in $C_j$, then $t_i^{2(mr+j)}$ has not yet been used in any triangle, so we can add $\{\widehat{c}_j^r, \widehat{d}_j^r, t_i^{2(mr+j)}\}$ to the triangle packing. On the other hand, if $v_i$ occurs negated in $C_j$ then $t_i^{2(mr+j)-1}$ has not yet been used in any triangle, so we can add $\{\widehat{c}_j^r, \widehat{d}_j^r, t_i^{2(mr+j)-1}\}$ to the triangle packing. In total $mn(n+1)+m(n+1)$ triangles are packed. □

**Lemma 19.** *If $G$ has a triangle packing of size $mn(n+1) + m(n+1)$, then $\phi$ satisfiable.*

*Proof.* Observe that for any $j$ and $r$, every triangle that contains $\widehat{c}_j^r$ also contains $\widehat{d}_j^r$ and vice versa. Furthermore, if we remove all the vertices $\widehat{c}_j^r$ and $\widehat{d}_j^r$ for every $j$ and $r$ from $G$ we obtain a disconnected graph with $n$ connected components, $G[T_i \cup V(P_i)]$ for every $i$. Thus, the only way to pack $mn(n+1) + m(n+1)$ triangles in $G$ is to pack $mn(n+1)$ triangles in each component $G[T_i \cup V(P_i)]$ and in addition make sure that every pair $(\widehat{c}_j^r, \widehat{d}_j^r)$ is used in some triangle in the packing.

The only way to pack $mn(n+1)$ triangles in a component $G[T_i \cup V(P_i)]$ is to use every second triangle of the form $\{t_i^l, p_i^l, p_i^{l+1}\}$, except possibly at one point where two triangles on this form are skipped. By the pigeon hole principle there is an $0 \le r \le n$ such that for every $i$, every second triangle of the form $\{t_i^{2mr+l}, p_i^{2mr+l}, p_i^{2mr+l+1}\}$ for $1 \le l \le 2m$ is used. We make an assignment to the variables of $\phi$ as follows. For every $i$ such that $\{t_i^{2mr+1}, p_i^{2mr+1}, p_i^{2mr+l+1}\}$ is used, $v_i$ is set to true, and otherwise $\{t_i^{2mr+2}, p_i^{2mr+2}, p_i^{2mr+3}\}$ is used in the packing and $v_i$ is set to false. We prove that this assignment satisfies $\phi$.

For every $j$, the pair $(\widehat{c}_j^r, \widehat{d}_j^r)$ is used in some triangle in the packing. This triangle either contains $t_i^{2(mr+j)}$ or $t_i^{2(mr+j)-1}$ for some $i$. If it contains $t_i^{2(mr+j)}$, then $v_i$ occurs positively in $C_j$. Furthermore, since the triangle packing contains every second triangle of the form $\{t_i^{2mr+l}, p_i^{2mr+l}, p_i^{2mr+l+1}\}$ for $1 \le l \le 2m$, it follows that the triangle packing contains $\{t_i^{2mr+1}, p_i^{2mr+1}, p_i^{2mr+l+1}\}$ and hence $v_i$ is set to true. By an identical argument, if the triangle containing the pair $(\widehat{c}_j^r, \widehat{d}_j^r)$ contains $t_i^{2(mr+j)-1}$ then $v_i$ occurs negated in $C_j$ and $v_i$ is set to false. This concludes the proof. □

We now modify the construction to work for PARTITION INTO TRIANGLES instead of TRIANGLE PACKING. Given the graph $G$ as constructed from $\phi$, we construct a graph $G'$ as follows. For every $1 \leq i \leq n$ and $1 \leq l \leq m(n+1)$ we make a clique $Q_i^l$ on four vertices. The vertices of $Q_i^l$ are all adjacent to $t_i^{2l}$ and to $t_i^{2l-1}$. For every $i < n$ and and $1 \leq l \leq m(n+1)$ we make all vertices of $Q_i^l$ adjacent to all vertices of $Q_{i+1}^l$. Suppose that $2n+2$ is $p$ modulo 3 for some $p \in \{0,1,2\}$. We remove $p$ vertices from $Q_n^l$ for every $l \leq m(n+1)$.

**Lemma 20.** *$G$ has a triangle packing of size $t$ if and only if $G'$ can be partitioned into triangles.*

*Proof.* In the forward direction, consider a triangle packing of size $t$ in $G$ as constructed in Lemma 18. We can assume that the triangle packing has this form, because by Lemma 19 we have that $\phi$ is satisfiable.

For every fixed $1 \leq l \leq m(n+1)$, we proceed as follows. We know that there exists an $i$ such that both $t_i^{2l}$ and $t_i^{2l-1}$ are used in the packing. For every $i' \neq i$, exactly one out of $t_{i'}^{2l}$ and $t_{i'}^{2l-1}$ is used in the packing. For each such $i'$, we make a triangle containing the unused vertex out of $t_{i'}^{2l}$ and $t_{i'}^{2l-1}$ and two vertices of $Q_{i'}^l$. Then we "clean up" $Q_1^l, \ldots, Q_n^l$ as follows.

In particular, we start with the yet unused vertices of $Q_1^l$. There are two of them. Make a triangle containing these two vertices and one vertex of $Q_2^l$. Now $Q_2^l$ has one unused vertex left. Make a triangle containing this vertex and the two unused vertices of $Q_3^l$. Continue in this fashion until arrive at $Q_i^l$. At this point we have used 0, 1 or 2 vertices of $Q_i^l$ a triangle containing some vertices in $Q_{i-1}^l$. The case when we have used 0 vertices of $Q_i^l$ also covers the case that $i=1$. If we only used 0 or 1 vertices of $Q_i^l$, then we add a triangle that contains 3 vertices of $Q_i^l$. If there are still unused vertices in $Q_i^l$, then their number is either 1 or 2. We make a triangle containing these vertices and 1 or 2 of the unused vertices of $Q_{i+1}^l$. Now we proceed to $Q_{i+1}^l$ and continue in this manner until we reach $Q_n^l$. Since the total number of vertices in $\bigcup_{j \leq n} Q_j^l$ is $4n - p$, we know that $2n - 2$ of these vertices are used for triangles with vertices of $G$, and $2n + 2 - p$ is divisible by 3 the process described above will partition all the unused vertices of $\bigcup_{j \leq n} Q_j^l$ into triangles.

In the reverse direction, we argue that in any partitioning of $G'$ into triangles, exactly $t$ triangles must lie entirely within $G$. In fact, we argue that for any $l \leq m(n+1)$ exactly $n-1$ vertices out of $\bigcup_{i \leq n}\{t_i^{2l}, t_i^{2l-1}\}$ are used in triangles containing vertices from $\bigcup_{i \leq n} Q_i^l$.

Pick $1 \leq j \leq m$ and $r$ such that $l = mr + j$. Exactly one out of $\bigcup_{i \leq n}\{t_i^{2l}, t_i^{2l-1}\}$ is in a triangle with $\widehat{c}_j^r$ and $\widehat{d}_j^r$. Furthermore, for each $i \leq n$ the vertex $p_i^{2l}$ must be in a triangle either containing $t_i^{2l}$ or $t_i^{2l}$. Hence, at most $n-1$ vertices out of $\bigcup_{i \leq n}\{t_i^{2l}, t_i^{2l-1}\}$ are used in triangles containing vertices from $\bigcup_{i \leq n} Q_i^l$. Furthermore, any triangle containing $t_i^{2l}$ or $t_i^{2l-1}\}$ must either contain $p_i^{2l}, \widehat{c}_j^r$ or some vertex in $\bigcup_{i \leq n} Q_i^l$. Hence exactly $n-1$ vertices out of $\bigcup_{i \leq n}\{t_i^{2l}, t_i^{2l-1}\}$ are used in triangles containing vertices from $\bigcup_{i \leq n} Q_i^l$. Thus in the packing, exactly $3t$ vertices in $G'$ are contained in triangles completely inside $G$, and hence $G$ has a triangle packing of size $t$. □

To complete the proof for PARTITION INTO TRIANGLES we need to bound the pathwidth of $G'$.

**Lemma 21.** $\mathbf{pw}(G') \leq n + 10$.

*Proof.* We give a search strategy for $G'$ that uses $n + 10$ searchers. The strategy consists of $m(n+1)$ rounds and each round has $n$ stages. In the beginning of round $l$, $1 \leq l \leq m(n+1)$, there are searchers $n$ searchers placed, one on each vertex $p_i^{2l-1}$ for every $i$. Let $r$ and $1 \leq j \leq m$ be integers such that $l = mr + j$. We place one searcher on $\widehat{c}_j^r$ and one on $\widehat{d}_j^r$. These two searchers will stay put throughout the duration of this round. In stage $i$ of round $l$ we place searchers on all vertices of $Q_i^l$ and $Q_{i+1}^l$. Then we place searchers on $t_i^{2l-1}, t_i^{2l}, p_i^{2l}$ and $p_i^{2l+1}$. At the end of stage $i$ we remove the searchers from $Q_i^l, t_i^{2l-1}$, $t_i^{2l}$ and $p_i^{2l}$. We then proceed to the next stage. At the end of the round we remove the searchers from $\widehat{c}_j^r$ and $\widehat{d}_j^r$. Notice that now, there are searchers on $p_i^{2l+1}$ for every $i$, and the next round can commence. □

Lemmata 18, 19, 20 and 21 prove Theorem 6.

# 9 Conclusion

We have showed that for a number of basic graph problems, the best known algorithms parameterized by treewidth are optimal in the sense that base of the exponential dependence on treewidth is best possible. Recall that for DOMINATING SET and PARTITION INTO TRIANGLES, this running time was obtained quite recently using the new technique of fast subset sum convolutions [27]. Thus it could have been a real possibility that the running time is improved for some other problems as well.

The results are proved under the Strong Exponential Time Hypothesis (SETH). While this hypothesis is relatively recent and might not be accepted by everyone, our results at least make a connection between rather specific graph problems and the very basic issue of better SAT algorithms. Our results suggest that one should not try to find better algorithms on bounded treewidth graphs for the problems considered in the paper: as this would disprove SETH, such an effort is better spent on trying to disprove SETH directly in the domain of satisfiability. Finally, we suggest the following open questions for future work:

- Can we prove similar tight lower bounds under the restriction that the graph is planar? Or is it possible to find improved algorithms on bounded treewidth planar graphs?
- Can we prove tight lower bounds for problems parameterized not by treewidth, but by something else? Naturally, one should look at problems where the algorithm or the the running time suggests that the best known algorithm is optimal. Possible candidates are the $\mathcal{O}(2^k)$ time algorithm for STEINER TREE with $k$ terminals [2], the $\mathcal{O}(2^k)$ time randomized algorithm for $k$-PATH [29], and the $\mathcal{O}(2^k)$ (resp., $\mathcal{O}(3^k)$) time algorithms for EDGE BIPARTIZATION (resp., ODD CYCLE TRANSVERSAL) [16, 22].
- For the $q$-COLORING problem, we were able to prove lower bounds parameterized by the feedback vertex set number. Can we prove such bounds for the other problems as well?